\documentclass[journal=jacsat,manuscript=article]{achemso}

\usepackage[version=3]{mhchem} 
\usepackage{verbatim}
\usepackage{graphicx}
\usepackage[labelfont=bf]{caption}
\usepackage[labelfont=bf,singlelinecheck=off,justification=raggedright]{subcaption}
\usepackage{subcaption}

\usepackage{siunitx}
\graphicspath{{./img}}
\usepackage{tabularx}
\usepackage{tabulary}
\usepackage{soul}
\usepackage{textcomp, gensymb}

\usepackage{amsmath}
\usepackage[table,xcdraw]{xcolor}
\usepackage[export]{adjustbox}
\usepackage{url}

\author{Reisel Millan}
\affiliation{Department of Materials Science and Engineering, Massachusetts Institute of Technology, Cambridge, MA 02139}
\alsoaffiliation{Universitat Politècnica de València}

\author{Estefanía Bello-Jurado}
\affiliation{Instituto de Tecnología Química, Universitat Politècnica de València-Consejo Superior de Investigaciones Científicas, Avenida de los Naranjos s/n, 46022 Valencia, Spain}

\author{Manuel Moliner}
\affiliation{Instituto de Tecnología Química, Universitat Politècnica de València-Consejo Superior de Investigaciones Científicas, Avenida de los Naranjos s/n, 46022 Valencia, Spain}

\author{Mercedes Boronat}
\email{boronat@itq.upv.es}
\affiliation{Instituto de Tecnología Química, Universitat Politècnica de València-Consejo Superior de Investigaciones Científicas, Avenida de los Naranjos s/n, 46022 Valencia, Spain}

\author{Rafael Gomez-Bombarelli}
\email{rafagb@mit.edu}
\affiliation{Department of Materials Science and Engineering, Massachusetts Institute of Technology, Cambridge, MA 02139}

\title[Machine-learning accelerated molecular dynamics of Cu-CHA]
      {Effect of framework composition and \ce{NH3} on the diffusion of \ce{Cu+} in Cu-CHA catalysts predicted by machine-learning accelerated molecular dynamics}

\keywords{NNP, diffusion, SCR-NOx}

\begin{document}
\sloppy

\begin{abstract}
Cu-exchanged zeolites rely on mobile solvated \ce{Cu+} cations for their catalytic activity, but the role of framework composition on transport is not fully understood. Ab initio molecular dynamics simulations can provide quantitative atomistic insight but are too computationally expensive to explore large length- and time-scales or diverse compositions. We report a machine-learning interatomic potential that accurately reproduces ab initio results and effectively generalizes to allow multi-nanosecond simulations of large supercells and diverse chemical compositions. Biased and unbiased simulations of \ce{[Cu(NH3)2]+} mobility show that aluminum pairing in eight-membered rings accelerates local hopping, and demonstrate that increased \ce{NH3} concentration enhances long-range diffusion. The probability of finding two \ce{[Cu(NH3)2]+} complexes in the same cage - key for SCR-NOx reaction  - increases with Cu content and Al content, but does not correlate with the long-range mobility of \ce{Cu+}. Supporting experimental evidence was obtained from reactivity tests of Cu-CHA catalysts with controlled chemical composition.

\end{abstract}



Copper-exchanged zeolites play a crucial role as redox catalysts for some environmentally
relevant processes, like the partial methane oxidation to methanol or the selective catalytic reduction of nitrogen oxides with ammonia (\ce{NH3-SCR-NOx}). In both cases, the small pore Cu-SSZ-13 zeolite with the CHA structure has been reported as an efficient catalyst. \cite{Sushkevich2017SelectiveMethanol, Dinh2019ContinuousZeolites,Dinh2021BreakingCatalysts, DelCampo2021ActivationMaterials, Ohyama2022RelationshipsSpectroscopies, Kwak2010ExcellentNH3, Paolucci2016CatalysisZeolites, Gao2017SelectiveCatalysis, Paolucci2017a, Peden2019Cu/ChabaziteControl, Vennestrm2022AdvancesNOx}

The \ce{NH3-SCR-NOx} reaction is currently employed for the removal of nitrogen oxides (NOx) from exhaust gases in diesel vehicles and stationary plants, through a redox catalytic cycle in which \ce{Cu+} is oxidized to \ce{Cu^2+} by \ce{NO2} or \ce{NO}+\ce{O2}, and then reduced to \ce{Cu+} by reaction of \ce{NH3} and \ce{NO} forming harmless \ce{N2 + H2O} (Scheme \ref{fig:scheme_intro}). \cite{Moreno-Gonzalez2019, Chen2020ACu-CHA, Borfecchia2018Cu-CHACatalysis, signorile2022review}
This understanding of the reaction mechanism has enabled development of optimized catalysts by tuning framework topology, composition and copper speciation. In the as-prepared catalysts, \ce{Cu+} and \ce{Cu^2+} cations are directly coordinated to the zeolite framework forming heterogeneous active sites, while under reaction conditions \ce{NH3} solvates the \ce{Cu+} cations forming mobile \ce{[Cu(NH3)2]+} complexes that act as dynamic active sites, resembling homogeneous catalysts but within the confinement of the zeolite pores. At low temperature (T $<$ 523 K) the oxidation step involves transient dimeric \ce{Cu+(NH3)2-O2-Cu+(NH3)2} species whose formation requires the simultaneous presence of two \ce{[Cu(NH3)2]+} monomers in the same \textit{cha} cage. The hops between adjacent \textit{cha} cages are modulated by size exclusion effects and also by the attractive interaction between the positively charged \ce{[Cu(NH3)2]+} complexes and the negatively charged framework Al sites\cite{Gao2017SelectiveCatalysis, Paolucci2017a, signorile2022review,Millan2020,Millan2021}. Thus, structural properties like Al content and distribution, Cu loading or Brønsted acid site density, as well as the interaction of the Cu active sites with the reactants, in particular \ce{NH3}, might affect the mobility of Cu cations and consequently the \ce{NH3-SCR-NOx} reaction rate. This has been evidenced by recent studies combining catalytic activity tests with \textit{operando} XAS or EPR spectroscopy, \cite{Martini2022AssessingSpectroscopy, Krishna2023InfluenceNH3,Marberger2018Time-resolvedCu-SSZ-13, Becher2020ChemicalSpectrotomography, Wu2023InterplayNH3-SCR} and \textit{ab initio} molecular dynamics (AIMD) simulations have been successfully applied to provide atomistic insight into the dynamic nature of the \ce{Cu+} cations under reaction conditions\cite{Paolucci2017a, Millan2020, Millan2021}.

\begin{scheme}[H]
\centering
\includegraphics[width=0.8\textwidth]{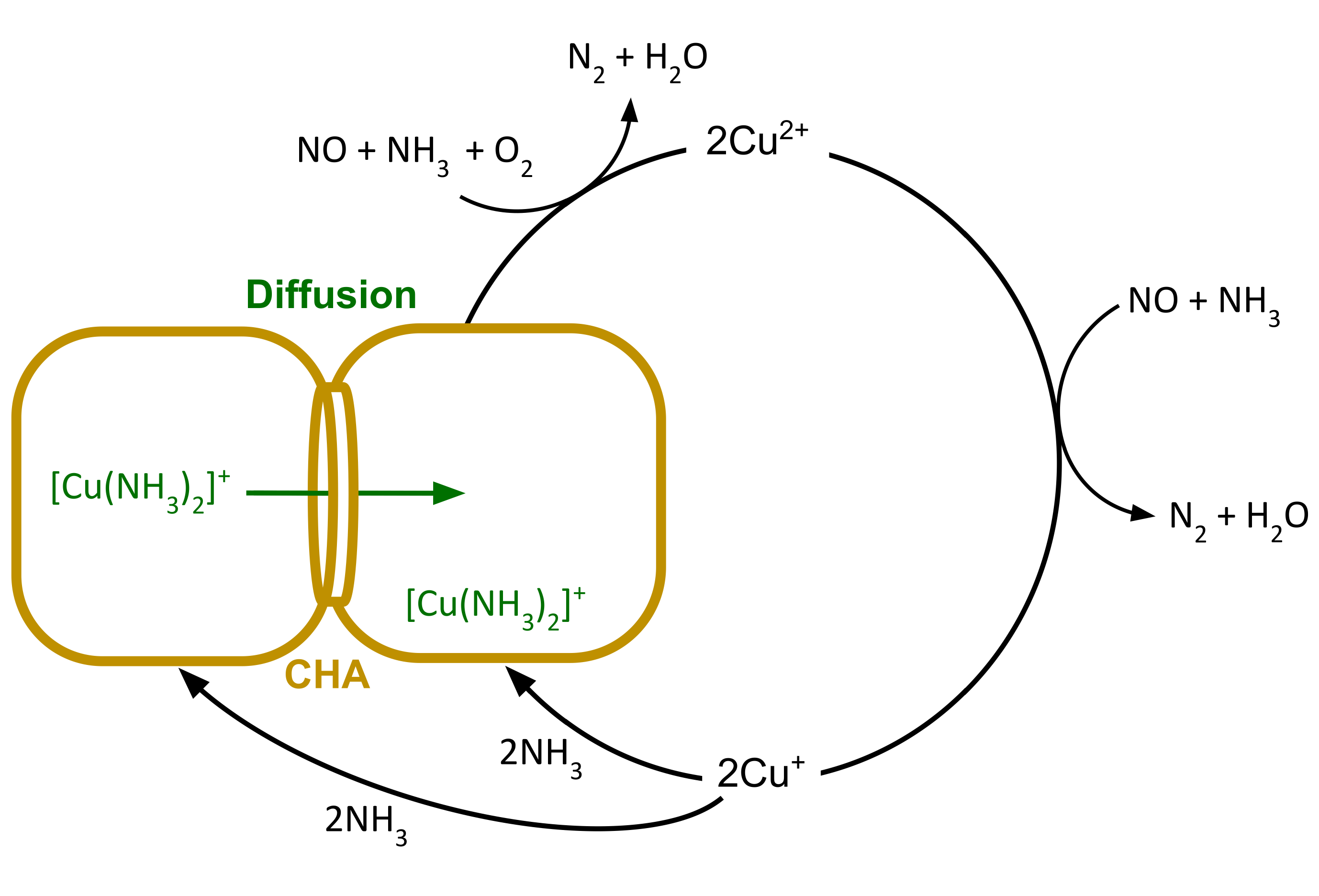}
\caption{Illustration of the low temperature SCR-NOx redox cycle.}
\label{fig:scheme_intro}
\end{scheme}

The cost of AIMD simulations limits their applicabiltiy to a few selected systems at a time, at small length- and time- scales. The timescale limitation may be partially bypassed with enhanced sampling methods, such as umbrella sampling (US), which have been used to study the slow diffusion of copper complexes in CHA.\cite{Paolucci2017a, Millan2021} However, the systematic exploration of parameters such as Si/Al ratio, Al distribution, Cu/Al ratio, \ce{NH3} concentration or the presence of Brønsted acid sites and compensating \ce{NH4+} cations has not been yet possible. 

Machine learning (ML) has demonstrated broad applicability in materials science\cite{Axelrod2022e,Schmidt2019a} and heterogeneous catalysis. \cite{Ma2020, Goldsmith2018, SchlexerLamoureux2019MachineCatalysis, Gu2020PracticalScreening} Machine learning potentials (MLPs), when trained with a sufficiently large and diverse dataset, can match the accuracy of quantum chemistry methods at a fraction of the computational cost\cite{Behler2011a, Behler2017, Botu2017, Mueller2020, VonLilienfeld2020,Noe2019, Jensen2019}. 
This allows the study of larger and more realistic systems and more complex scientific problems \cite{Butler2018, Schmidt2019a, Erlebach2022AccuratePotentials, Moliner2019a}, in particular those requiring the use of molecular dynamics simulations \cite{Chmiela2018, Sosso2012NeuralGeTe, Bernstein2019QuantifyingSilicon}.
A broad variety of MLPs based on neural networks - so called  neural network potentials (NNPs) - have been developed in the last years:  ANI \cite{Smith2017,Smith2017b,Smith2019,Devereux2020}, Deep Tensor Neural Networks \cite{Schutt2017}, SchNet\cite{Schutt2019}, DeepPotentialNet \cite{Zhang2018a}, MEGNet\cite{Chen2019GraphCrystals}, DimeNet\cite{Klicpera2020a}, OrbNet\cite{Qiao2020}, PaiNN\cite{Schutt2021}, NequIP \cite{Batzner2021E3-EquivariantPotentials}, and  successfully used to study solids systems\cite{Artrith2016, Erlebach2022AccuratePotentials, Vandenhaute2023MachineApproach, Bocus2023NuclearDynamics}, ion diffusion \cite{winter2022simulations}, and chemical reactions \cite{Gastegger2015, ang2021active, du2023machinelearningaccelerated}.

Here, we leveraged these innovations and trained an NNP capable of describing \ce{[Cu(NH3)2]+} species in aluminosilicate CHA with varying composition and \ce{NH3} concentration. The trained NNP proved accurate and transferable, and acquiring all the training data was less costly than one traditional AIMD simulation. Biased MD simulations reproduced free energy profiles from DFT, and provided insight into transport for over a dozen combinations of Al distribution and \ce{NH4+} presence. Unbiased MD simulations were scaled to thousands of atoms for nanoseconds, and achieved a more realistic representation of the importance of Al density and distribution, Cu loading, and adsorbed \ce{NH3} on the mobility of \ce{Cu+} cations in Cu-CHA catalysts. 

These results show that the activation free energy for \ce{[Cu(NH3)2]+} hops between adjacent cages is lower for windows containing Al pairs, but also that this is a local effect with only weak influence on long-range mobility. \ce{[Cu(NH3)2]+} migration to remote cages requires the simultaneous displacement of charge-compensating \ce{NH4+} which show a lower mobility that is enhanced by excess \ce{NH3}. Lastly, simulations with large supercells show that the probability of finding two \ce{[Cu(NH3)2]+} complexes in the same cage, a pre-requisite for the SCR-NOx reaction, increases with Cu loading, and also with the Al content in the zeolite. We confirm these trends experimentally through catalytic tests of  Cu-CHA samples with controlled Si/Al and Cu/Al ratios. 

\section{Results}
\subsection{Neural network potential} 
\begin{figure}[H]
\raggedleft
\begin{subfigure}[b]{\textwidth}
    \includegraphics[width=\textwidth]{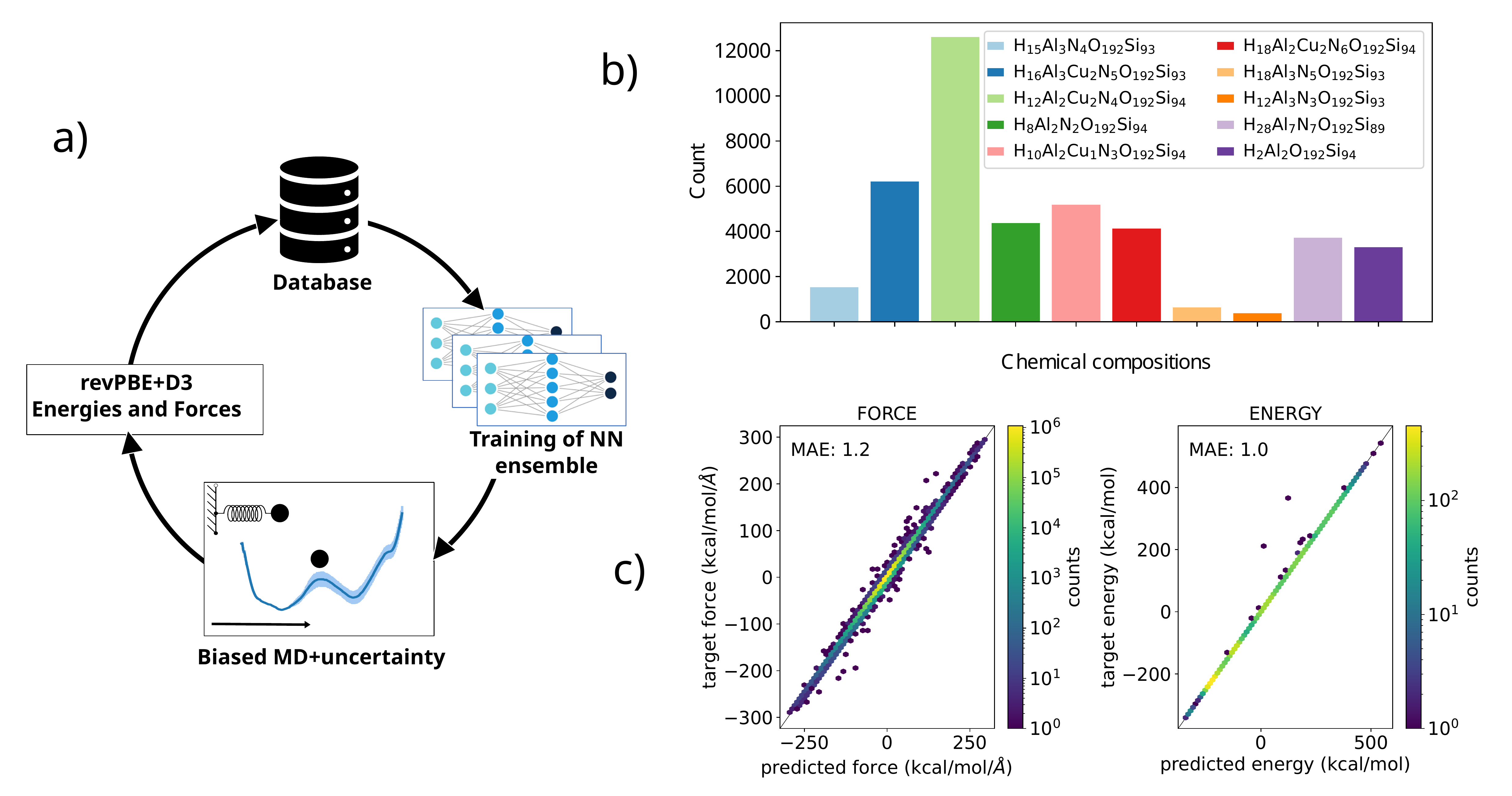}
\end{subfigure}
\caption{\textbf{Neural Network Potential.} (a) Illustration of the active learning cycle. (b) Distribution of the chemical composition of the dataset. (c) Correlation between the predicted and target energies (left) and  forces (right) of the last generation potential. }
\label{fig:fig1}
\end{figure}

NNPs are highly accurate but they struggle to extrapolate outside their training data. In order to ensure robust and accurate production simulations, our NNP was trained on data gathered through multiple generations of active learning (AL) using a query by-committee approach \cite{Behler2015, smith2018less, Schran2020, Musil2019, Peterson2017b, Lookman2019, Shapeev2020a, Imbalzano2021, wang2020active}. A committee (ensemble) of NNPs was trained on the available labeled data at each iteration, and new data was collected based on the disagreement (variance) of the prediction of the committee members on newly generated geometries, as described in Computational Details Section and illustrated in Figure \ref{fig:fig1}a. The last generation of the NNP was trained on a complete dataset containing 42k revPBE+D3 force calculations on structural models with a diverse set of atomic local environments, ranging from 290 to 323 atoms per supercell, summarized in Table \ref{tab:dataset_composition_training}  (see structural models in Figure \ref{fig:snapshots_cell_sizes}). The chemical compositions in the dataset (Figure \ref{fig:fig1}b) cover a range of Si/Al ratio from 47 to 13. All DFT calculations in the training data were electrically neutral with negative charges arising from Al substitution being compensated by \ce{[Cu(NH3)2]+}, \ce{NH4+}, or \ce{H+}\cite{wang2020differentiable,Schwalbe-Koda2021,tan2023singlemodel}. 

The active learning strategy was capable of automatically adding new, diverse, and informative chemical environments to the training pool at each of the pre-selected compositions through a combination of MD and uncertainty quantification. It generated informative training data for a number of chemical processes that occur during the reaction, but were not present in the initial training data. These include adsorption and protonation of \ce{NH3} on the Brønsted acid sites to form \ce{NH4+} cations, exchange between a gas phase \ce{NH3} molecule and one of the two \ce{NH3} ligands of the \ce{[Cu(NH3)2]+} complex, and proton transfer from \ce{NH4+} to \ce{NH3}. The diffusion  of \ce{[Cu(NH3)2]+} complexes
through the 8R windows that connect adjacent \textit{cha} cages has a higher activation barrier. Therefore representative training data was obtained through the same enhanced sampling approach as the production simulations (Figure \ref{fig:cv}).

This strategic combination of biased MD with uncertainty quantification allowed efficient sampling of the relevant regions on the PES with a small and diverse number of DFT evaluations. Figure \ref{fig:features} illustrates the structural diversity in the final dataset by means of a 2D projection of the local chemical environments around each Al atom in our data using UMAP\cite{McInnes2018} on the feature vectors learned by the NNP \cite{xie2018}. Atoms with similar local environments have similar feature vectors and appear close on the UMAP plot. The overlap among the chemical compositions suggests a nearly continuous sampling of the Al local environment. 

Figure \ref{fig:fig1}d shows the correlation between predicted and target energies and forces for a held-out test set. The mean absolute error of the predicted energies and forces are 0.98 kcal/mol and 1.2 kcal/mol/\text{\AA} respectively, indicating  that the NNP is capable of predicting the energies and forces with chemical accuracy.

\subsection{Effect of Al distribution on \ce{[Cu(NH3)2]+} diffusion through 8R windows from biased simulations}
\begin{figure}[H]
\captionsetup[subfigure]{labelformat=empty, justification=centering}
        \begin{subfigure}[b]{\textwidth}
            \centering
            \includegraphics[width=\textwidth]{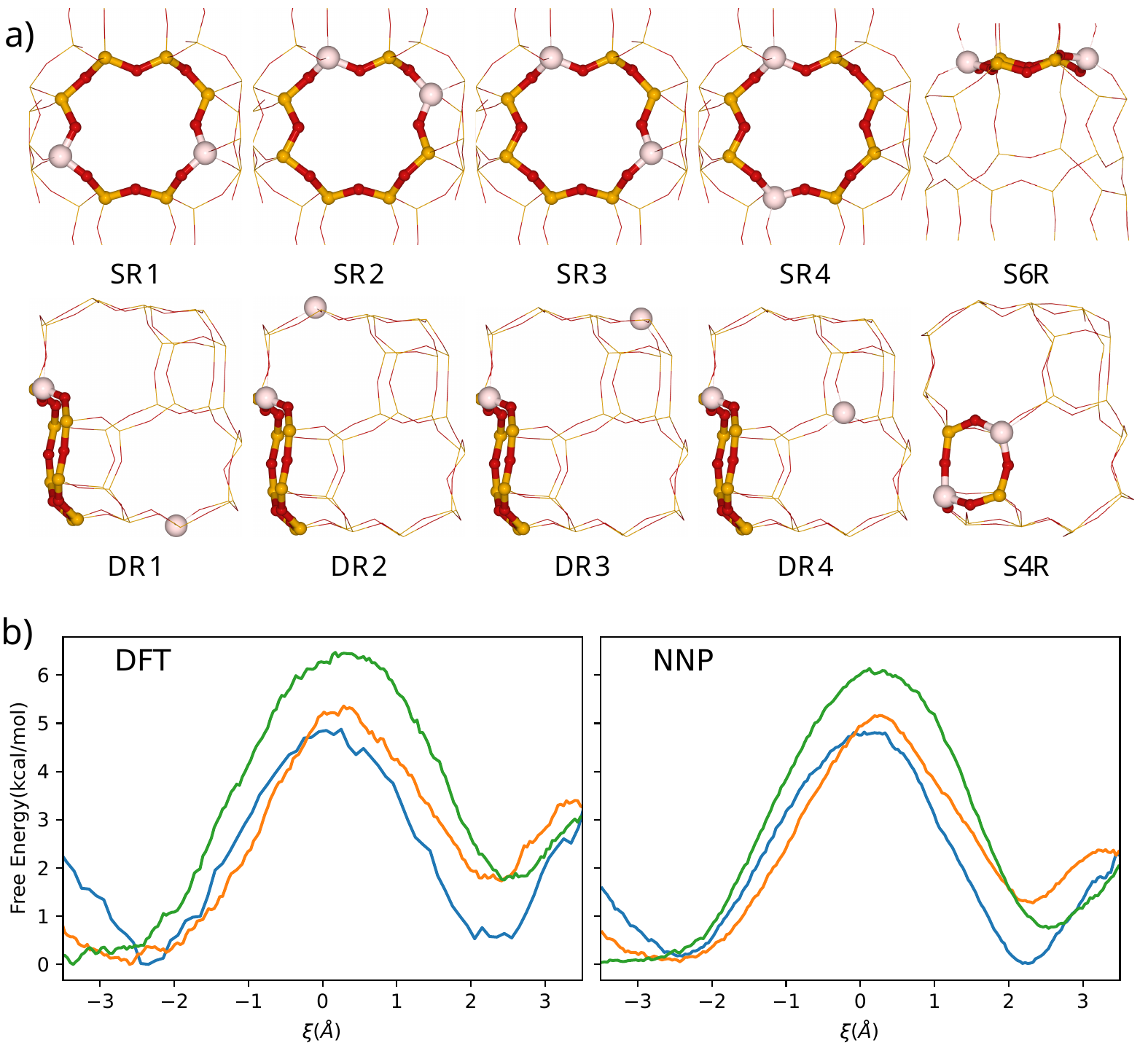}
        \end{subfigure}
    
    \caption{\textbf{Structural models and accuracy of NNP biased simulations.} (a) Representation of the different Al pair distributions used in the biased simulations. The atoms in the ring containing the Al pair are highlighted. Si, O and Al atoms are depicted in orange, red and light brown, respectively. (b) Comparison of free energy profiles for \ce{[Cu(NH3)2]+} diffusion through the 8R window of SR1 (orange), SR4 (blue) and DR2 (green) systems obtained from DFT (left) and NNP (right) based biased simulations at 423 K.}
    \label{fig:snapshots_of_alpairs_and_nn_vs_dft}
\end{figure}

The favorable speed of the NNP accelerates US MD simulations by orders of magnitude over DFT, and enabled systematic exploration of the role of Al distribution in well converged simulations. Ten different structural models with \ce{H12Al2Cu2N4O192Si94} composition were built (Fig \ref{fig:snapshots_of_alpairs_and_nn_vs_dft}a), each containing two framework Al atoms compensated with two \ce{[Cu(NH3)2]+}, and distributed either in the same 8R (SR1, SR2, SR3 and SR4), in different 8Rs (DR1, DR2, DR3 and DR4), in the same 4R (S4R) or in the same 6R (S6R).   

Because the error of the predicted forces is not a sufficient metric for true performance in production simulations \cite{fu2022forces}, the NNPs were further validated by comparing NNP and DFT free energy profiles for \ce{[Cu(NH3)2]+} diffusion through the 8R window of DR2, SR1 and SR4. Due to the high computational cost of producing reference DFT biased simulations, a smaller hexagonal 126-atom unit cell was used. The free energy profiles for SR1, SR4 and DR2 (Figure \ref{fig:snapshots_of_alpairs_and_nn_vs_dft}b) and the calculated activation free energies  ($\Delta G_\mathrm{act}$),  5.2 and 5.3 kcal/mol for SR1, 4.8 and 4.8 kcal/mol for SR4, and  6.1 and 6.4 kcal/mol for DR2, are in excellent agreement at both computational levels.

The smoother NNP traces are a consequence of more abundant sampling (4.8 ns total, compared to 0.8 for DFT) given the advantageous computational cost of NNP (20 ps of AIMD required over a week on CPU cores (Intel(R) Xeon(R) CPU E5-2650 v3 @ 2.30GHz), compared to 20 min on a Tesla V100-32 GB GPU for NNP).

\begin{figure}[H]
\captionsetup[subfigure]{labelformat=empty, justification=centering}
    \centering
    
    \begin{subfigure}[b]{\textwidth}
        \includegraphics[width=\textwidth]{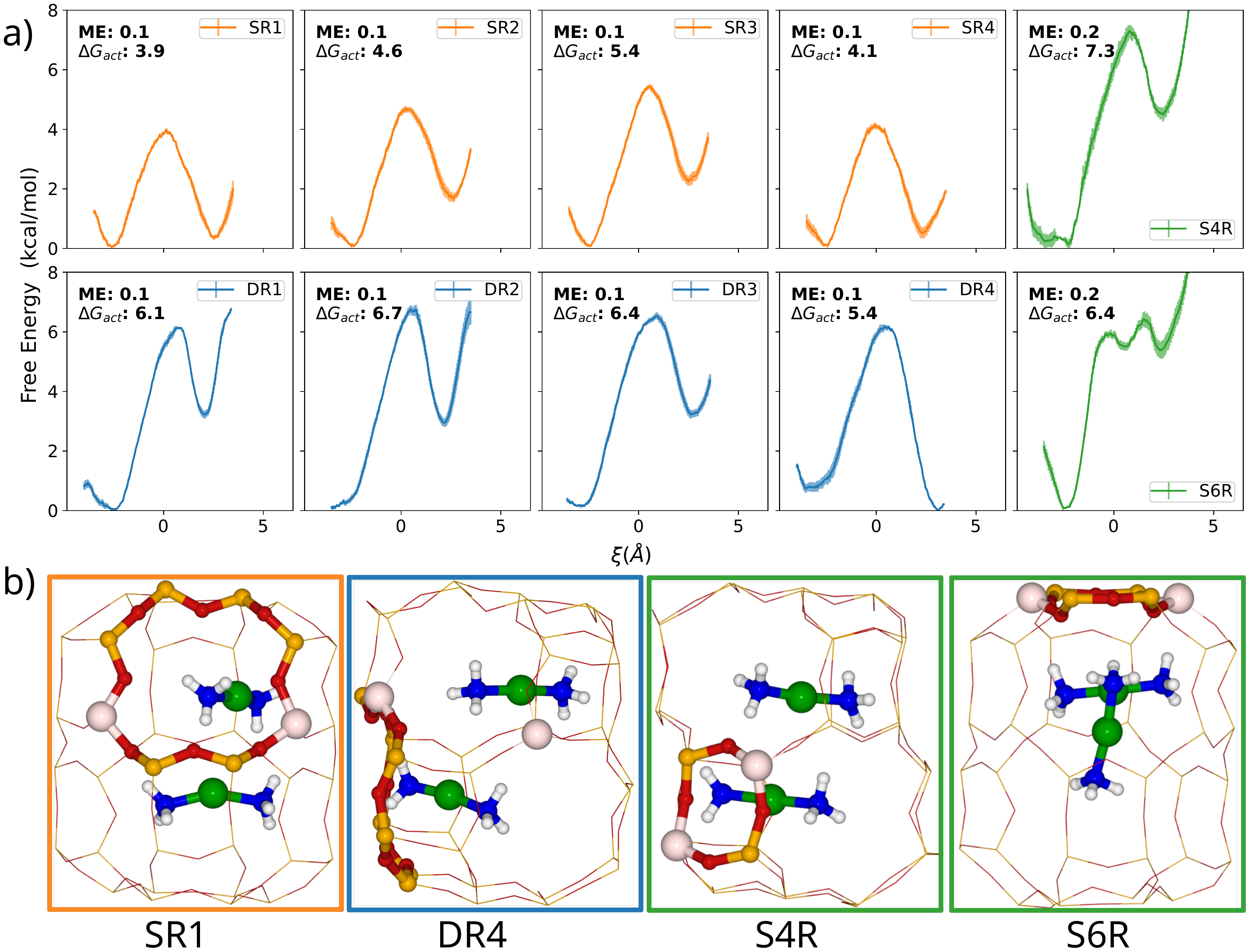}
    \end{subfigure}
    
    \caption{\textbf{NNP biased simulations of \ce{[Cu(NH3)2]+} diffusion}. (a) Free energy profiles for the diffusion of one \ce{[Cu(NH3)2]+} complex from a cage A to a neighboring cage B occupied by another \ce{[Cu(NH3)2]+} complex and (b) snapshots of the final state at $\xi$ = 2.5 corresponding to the diffusion of \ce{[Cu(NH3)2]+} through the 8R windows of SR1, DR4, S4R and S6R systems at 423 K.}
    \label{fig:al_effect}
\end{figure}

Then, the free energy profiles for \ce{[Cu(NH3)2]+} diffusion between neighboring cages in the ten systems depicted in Figure \ref{fig:snapshots_of_alpairs_and_nn_vs_dft}a were obtained from NNP biased simulations at 423 K using the larger \ce{H12Al2Cu2N4O192Si94} models. The profiles are plotted in Figure \ref{fig:al_effect} and the corresponding values of activation ($\Delta G_\mathrm{act}$) and reaction ($\Delta G$) free energies are summarized in Table \ref{tab:free_energies_us_al_pairs}. The shaded area in each profile shows the standard error calculated from three independent simulations using three different NNPs trained on the same dataset. The average value of the standard error, 0.1 kcal/mol in all cases, indicates a low uncertainty in the prediction of the free energy and well-converged statistics.

For the systems with two Al atoms in the same 8R (orange profiles in Figure \ref{fig:al_effect}), $\Delta G_\mathrm{act}$ values range from 3.9 to 5.4 kcal/mol and the reaction is slightly endergonic with $\Delta G$ values between 0.4 and 2.3 kcal/mol. In all other cases, $\Delta G_\mathrm{act}$ are higher than 6 kcal/mol and $\Delta G$ are larger than 3 kcal/mol, with the only exception of the DR4 system for which the process is slightly exergonic. The Al distribution in the DR4 model is the same as in SR1, but the diffusion occurs through different 8R windows (see snapshots in Figure \ref{fig:al_effect}).  In both cases, the stability of the final state with the two \ce{[Cu(NH3)2]+} in the same cage is similar to that of the initial state with the two complexes in different cages, which suggests that this particular Al distribution might favor the formation of the \ce{[Cu(NH3)2]+-OO-[Cu(NH3)2]+} dimers involved in the low-temperature \ce{NH3}-SCR-NOx reaction. In contrast, distributions with two Al in the same 4R or 6R hinder the formation of such dimeric intermediates, because only one of the  two \ce{[Cu(NH3)2]+} can stay near the Al atoms (see snapshots in Figure \ref{fig:al_effect}) while the second \ce{[Cu(NH3)2]+} is forced to remain in too close contact with the first, resulting in its diffusion through a different 8R to another empty cage.
These results demonstrate that Al distribution affects the movement of  \ce{[Cu(NH3)2]+} species between cages and the stability of pairs of \ce{[Cu(NH3)2]+} complexes in the same cage, and points to a positive effect of Al pairs in 8R on the rate of the low-temperature \ce{NH3}-SCR-NOx reaction.

\subsection{Effect of \ce{NH4+} on the diffusion of \ce{[Cu(NH3)2]+} through 8R windows from biased simulations}

\begin{figure}[H]
\captionsetup[subfigure]{labelformat=empty, justification=centering}
    \centering
\begin{minipage}[c]{0.9\textwidth}
    \begin{subfigure}[c]{.499\textwidth}
    \centering
   \includegraphics[width=\textwidth]{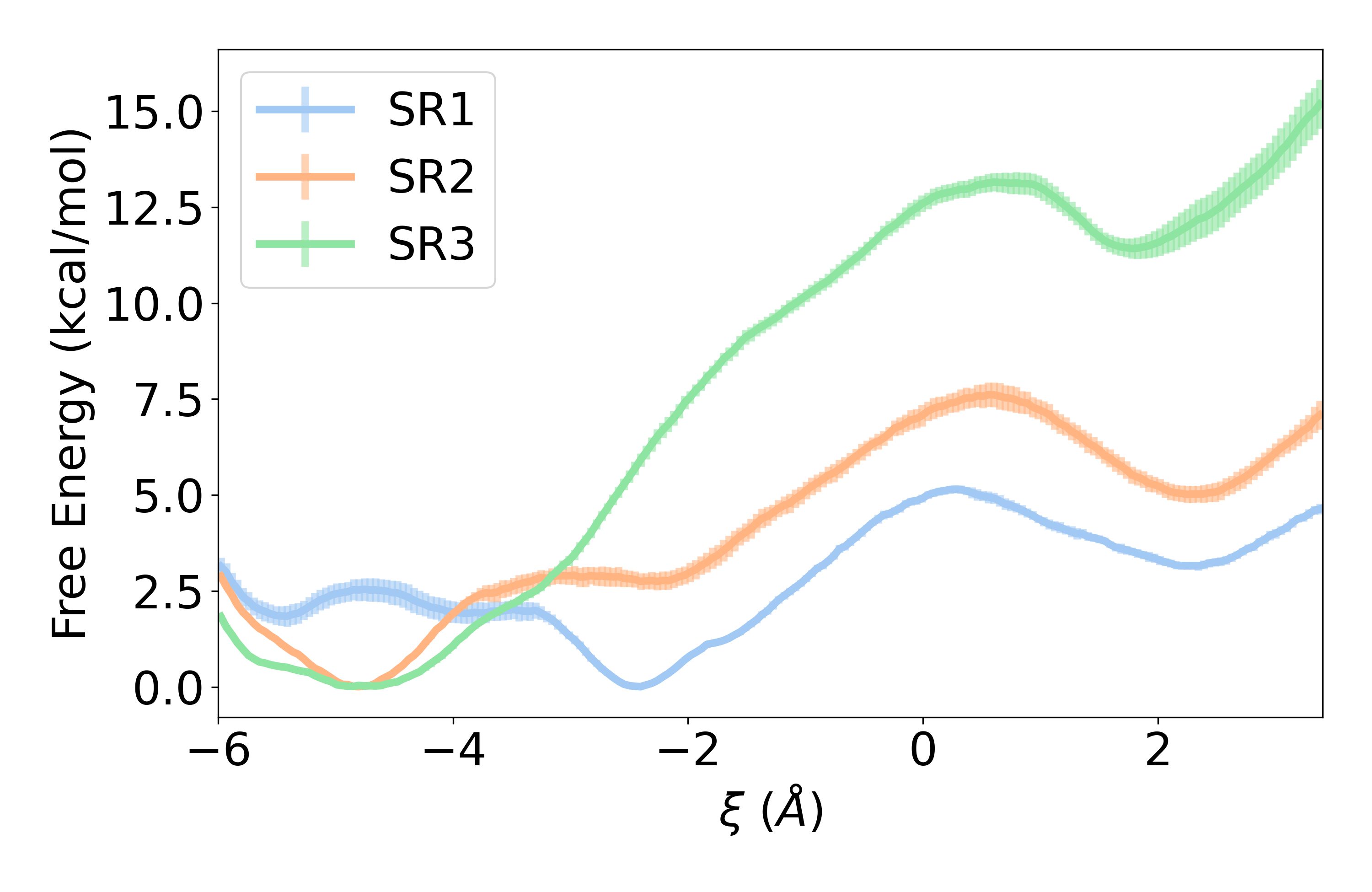}
    \end{subfigure}%
    \hfill
    \begin{subfigure}[c]{.249\textwidth}
    \centering
   \includegraphics[width=\textwidth]{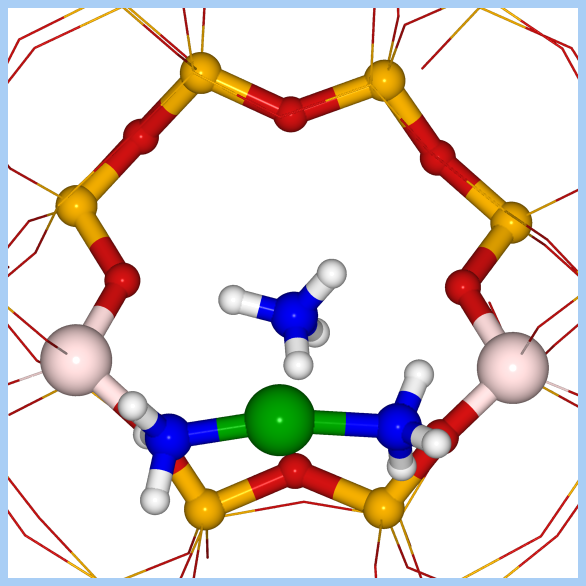}
   \caption{SR1 ($\xi$ = -2.5)}
    \end{subfigure}%
    \hfill
    \begin{subfigure}[c]{.249\textwidth}
    \centering
   \includegraphics[width=\textwidth]{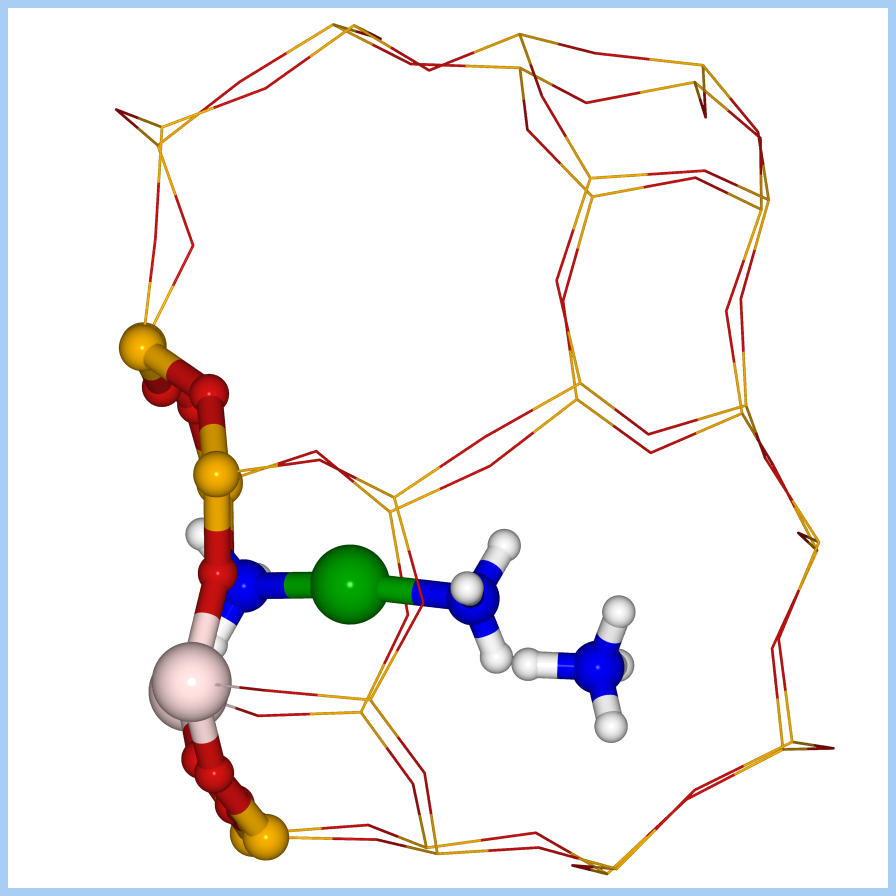}
      \caption{SR1 ($\xi$ = 2.5)}
    \end{subfigure}%
    \hfill
    \begin{subfigure}[c]{.249\textwidth}
    \centering
   \includegraphics[width=\textwidth]{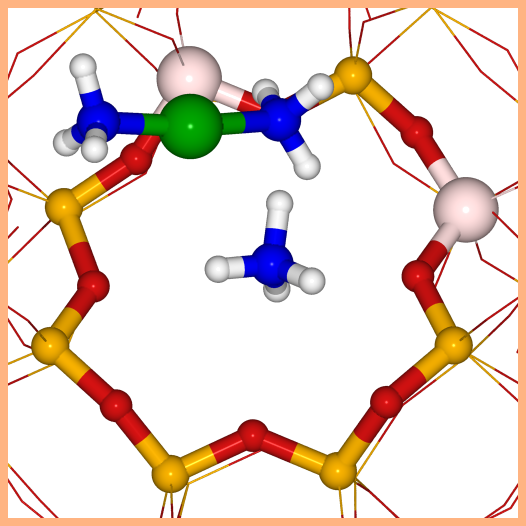}
      \caption{SR2 ($\xi$ = -4.8)}
    \end{subfigure}%
    \hfill
    \begin{subfigure}[c]{.249\textwidth}
    \centering
   \includegraphics[width=\textwidth]{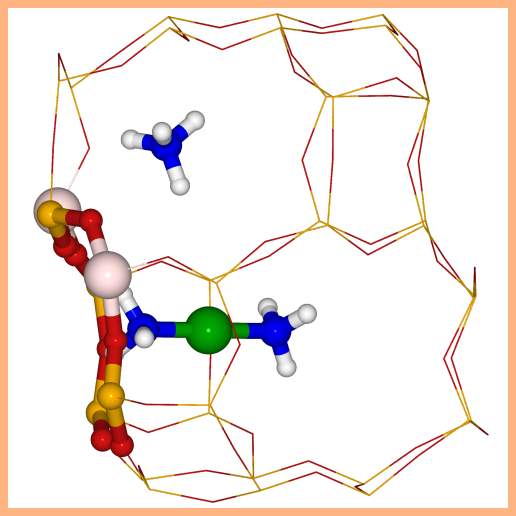}
      \caption{SR2 ($\xi$ = 2.5)}
    \end{subfigure}%
    \hfill
    \begin{subfigure}[c]{.249\textwidth}
    \centering
   \includegraphics[width=\textwidth]{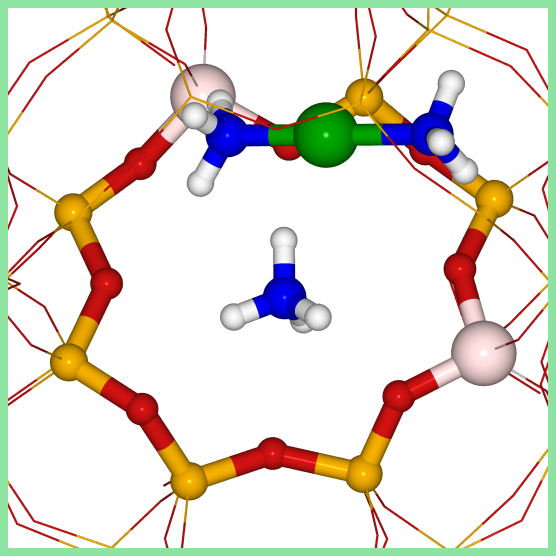}
      \caption{SR3 ($\xi$ = -4.8)}
    \end{subfigure}%
    \hfill
    \begin{subfigure}[c]{.249\textwidth}
    \centering
   \includegraphics[width=\textwidth]{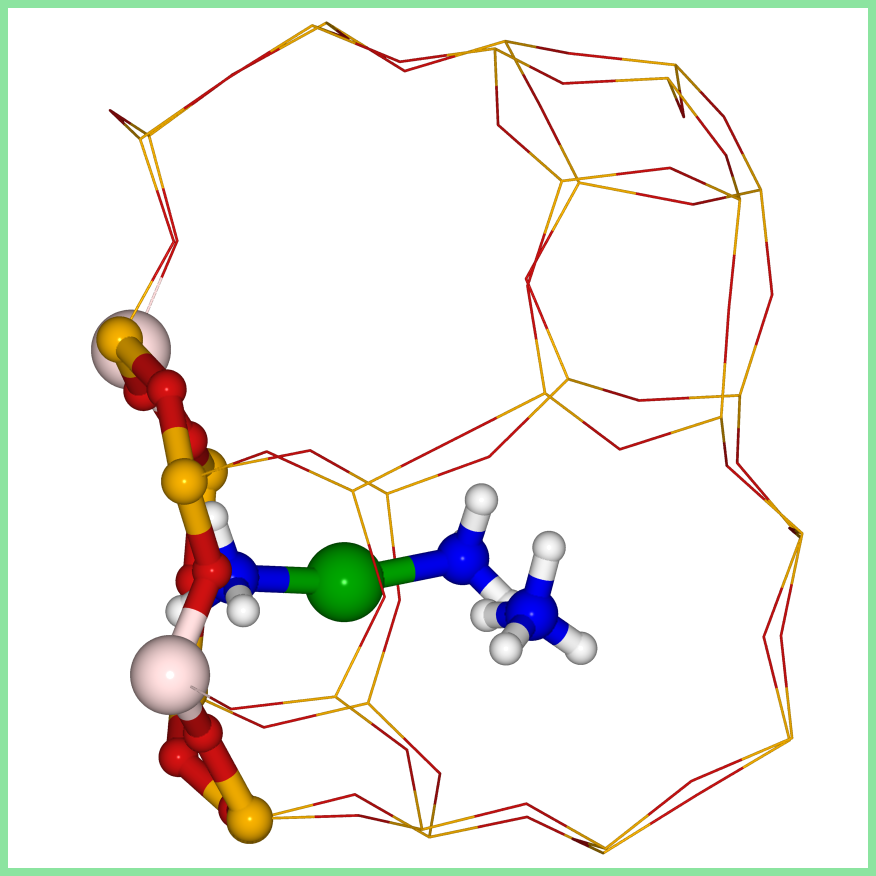}
      \caption{SR3 ($\xi$ = 2.0)}
    \end{subfigure}
\end{minipage}
\caption{\textbf{Effect of \ce{NH4+} on the diffusion of \ce{[Cu(NH3)2]+} complex.} Free energy profiles  for \ce{[Cu(NH3)2]+} diffusion between neighboring cages in the presence of \ce{NH4+} in three different systems obtained from NNP-based biased simulations at 423 K, and snapshots of the systems SR1, SR2 and SR3 at the initial and final minimum states. Si, O, Al, H, Cu and N atoms are depicted in orange, red, light brown, white, green and blue, respectively. }
\label{fig:diffusion_nh4_blocking}
\end{figure}

The mobility of \ce{[Cu(NH3)2]+} complexes within the zeolite microporous structure is affected by the presence of other molecules involved in the reaction, among which \ce{NH3} is the most abundant and the one with the largest impact on diffusion and reactivity.\cite{Millan2020, Millan2021,Marberger2018Time-resolvedCu-SSZ-13,Becher2020ChemicalSpectrotomography, Shwan2015Solid-StateTemperature, Vennestrm2019TheZeolites} 
Under reaction conditions \ce{NH3} is readily protonated on the Brønsted acid sites forming \ce{NH4+} cations that remain coordinated to the framework \ce{AlO4-} units and might partly block  the diffusion of \ce{[Cu(NH3)2]+} complexes through the 8R windows. To analyze this possibility we first performed NNP biased simulations at 423 K using the previously described \ce{H10Al2Cu1N3O192Si94} models with two framework Al atoms compensated now with one \ce{[Cu(NH3)2]+} complex and one \ce{NH4+} cation. The free energy profiles obtained for the three Al distributions considered (SR1, SR2 and SR3), plotted  in Figure \ref{fig:diffusion_nh4_blocking} are clearly different from those depicted in Figure \ref{fig:al_effect}, and the calculated activation free energies for \ce{[Cu(NH3)2]+} diffusion are also reflective of how the strong coordination of \ce{NH4+} modifies transport. 

Whereas in the previous simulations without \ce{NH4+} the most stable minimum for the initial state occurs at $\xi$ $\sim$ -2.5 \text{\AA}, with \ce{[Cu(NH3)2]+}  relatively close to the 8R, that is the case only for SR1 in the presence of \ce{NH4+}. In SR2 and SR3 models the most stable minimum lies at $\xi$ $\sim$ -4.8 \text{\AA}, with the \ce{[Cu(NH3)2]+} complex closer to the center of the cavity and farther away from the 8R to be crossed and therefore from the negatively charged \ce{AlO4-} units. The calculated $\Delta G_\mathrm{act}$, 5.2, 7.6 and 13.1 kcal/mol for SR1, SR2 and SR3, respectively, and $\Delta G$ values,  3.2, 5.0 and 11.4 kcal/mol for SR1, SR2 and SR3, respectively, are higher than those obtained for the corresponding systems in the absence of \ce{NH4+}. The snapshots of the final state at $\xi$ $\sim$ 2.5 \text{\AA} for SR1 and SR2 in Figure \ref{fig:diffusion_nh4_blocking} show that the \ce{NH4+} cation has been displaced from its initial position in the plane of the 8R to a position relatively close to one of the \ce{AlO4-} units. In SR3, however, the \ce{NH4+} cation has been displaced to the opposite side of the cage, far from the two \ce{AlO4-} sites present in the model, which would explain the instability of the system. The deviation across replicate profiles is wider in some regions, which we attribute to the fact that our simulations occasionally, but not exhaustively, sample the spontaneous reversible deprotonation of \ce{NH4+} cations to form \ce{NH3} and a Brønsted acid site, which is typically not accessible to traditional simulations.

Altogether, the results from the biased simulations suggest a potential blocking effect of \ce{NH4+} cations. However, their own mobility, as either \ce{NH4+} or \ce{NH3} following proton transfer to deprotonated \ce{AlO4-}, and possible migration from the 8R towards other nearby \ce{AlO4-} units that are not present in this model might modify this conclusion. The NNPs developed here open the possibility of running long-time unbiased MD simulations on larger systems with more realistic chemical compositions, allowing to capture the dynamics of \ce{[Cu(NH3)2]+}  and \ce{NH4+} globally and to observe long-range diffusion of both cationic species.

\subsection{Long-range diffusion of \ce{NH4+} and \ce{[Cu(NH3)2]+} species from unbiased MD simulations}
\begin{figure}[H]
\captionsetup[subfigure]{labelformat=empty, justification=centering}
\centering
\begin{subfigure}[b]{.24\textwidth}
\centering
    \includegraphics[width=\textwidth]{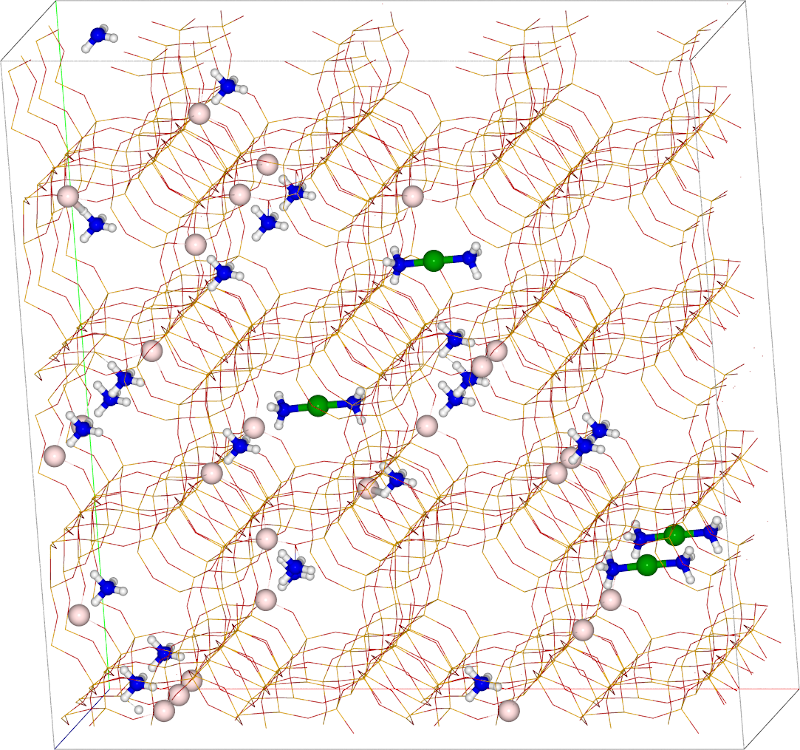}
    \caption*{L4}
\end{subfigure}%
\begin{subfigure}[b]{.24\textwidth}
\centering
    \includegraphics[width=\textwidth]{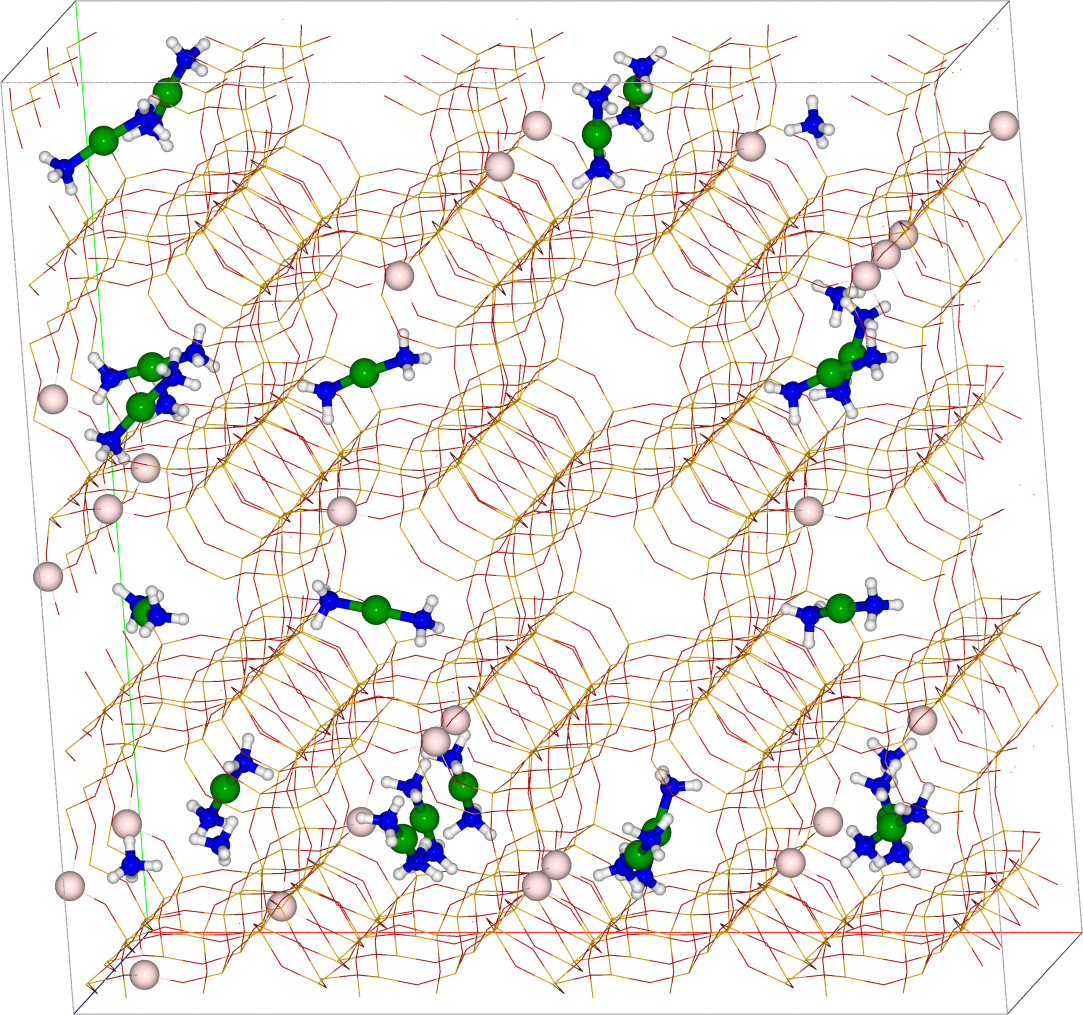}
    \caption*{L20}
\end{subfigure}
\begin{subfigure}[b]{.24\textwidth}
\centering
    \includegraphics[width=\textwidth]{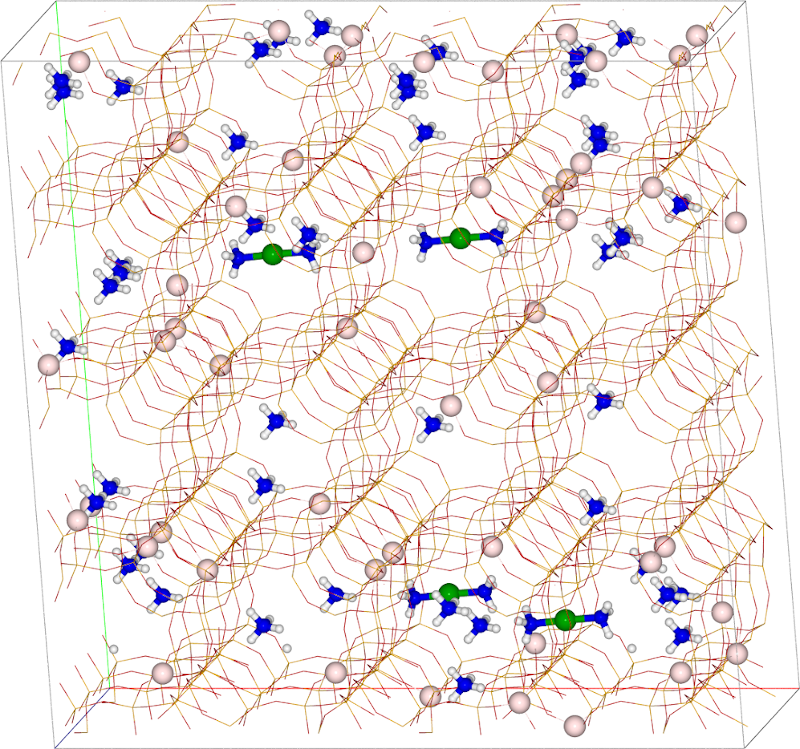}
    \caption*{M4}
\end{subfigure}
\begin{subfigure}[b]{.24\textwidth}
\centering
    \includegraphics[width=\textwidth]{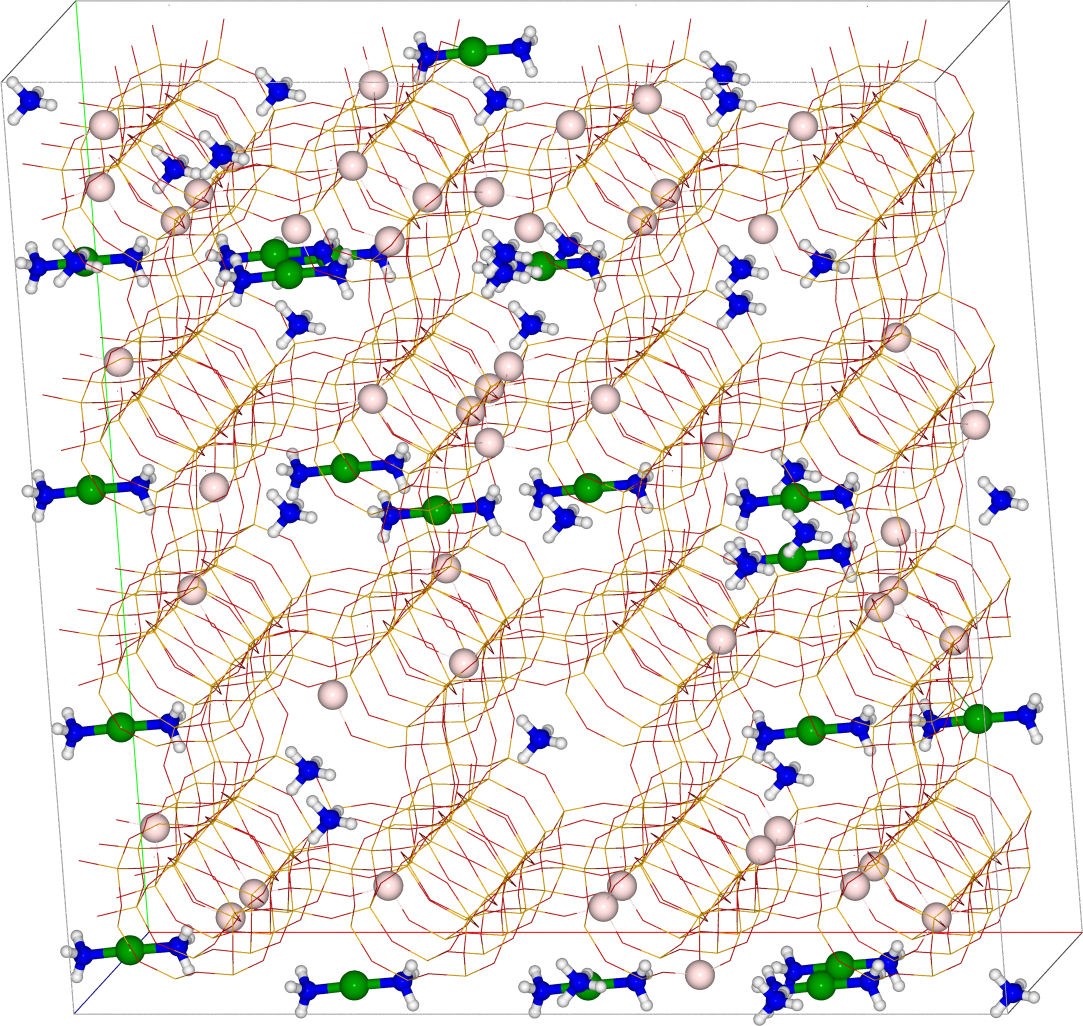}
    \caption*{M20}
\end{subfigure}
\begin{subfigure}[b]{.24\textwidth}
\centering
    \includegraphics[width=\textwidth]{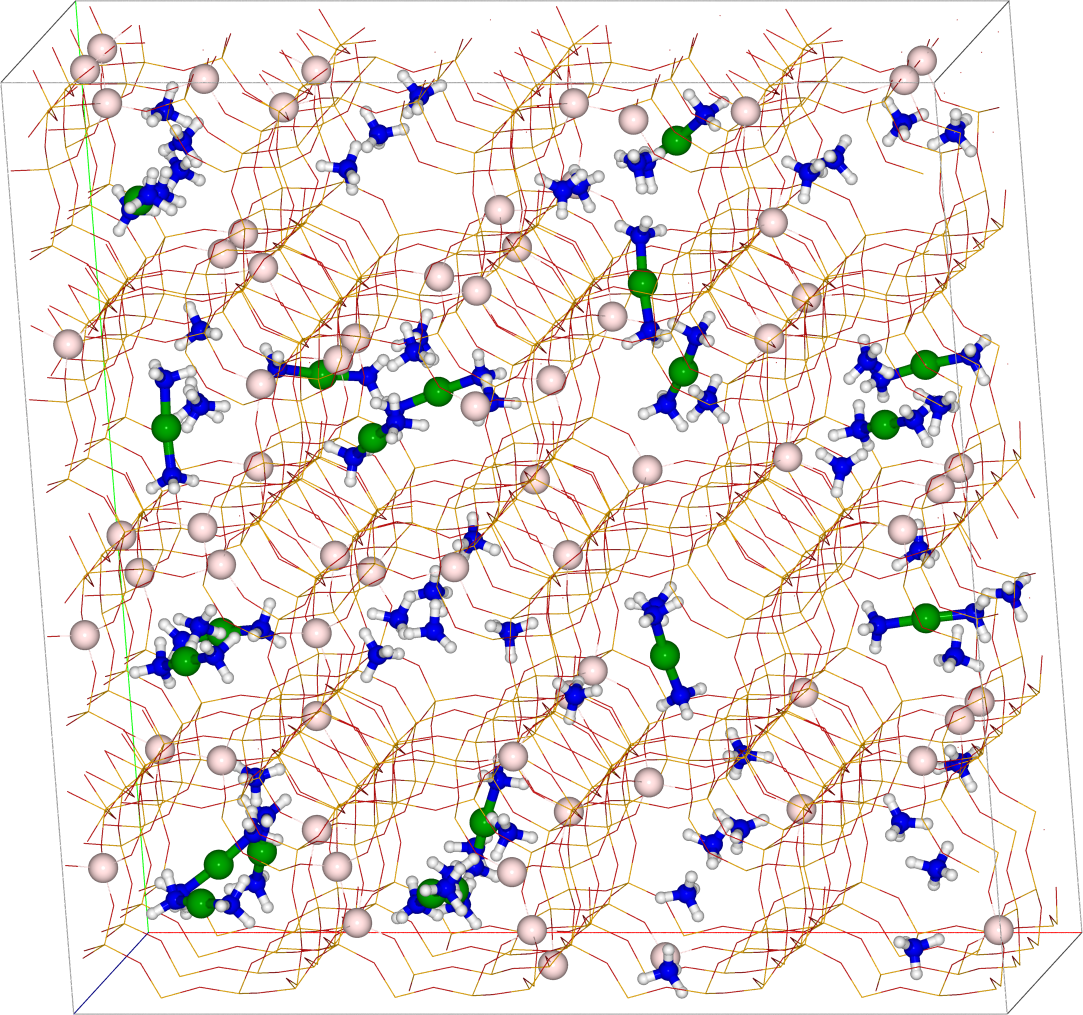}
    \caption{H6R}
\end{subfigure}
\begin{subfigure}[b]{.24\textwidth}
\centering
    \includegraphics[width=\textwidth]{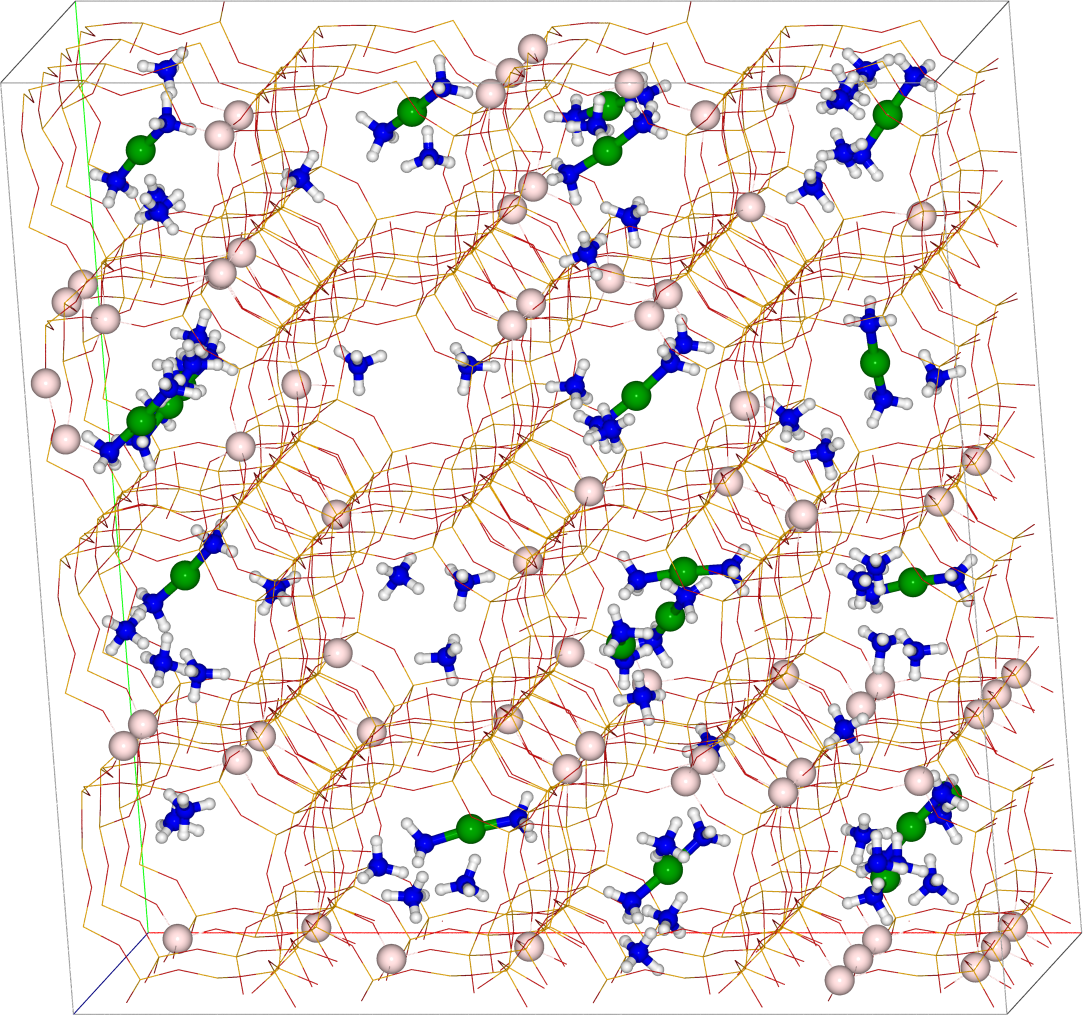}
    \caption{H8R}
\end{subfigure}
\begin{subfigure}[b]{.24\textwidth}
\centering
    \includegraphics[width=\textwidth]{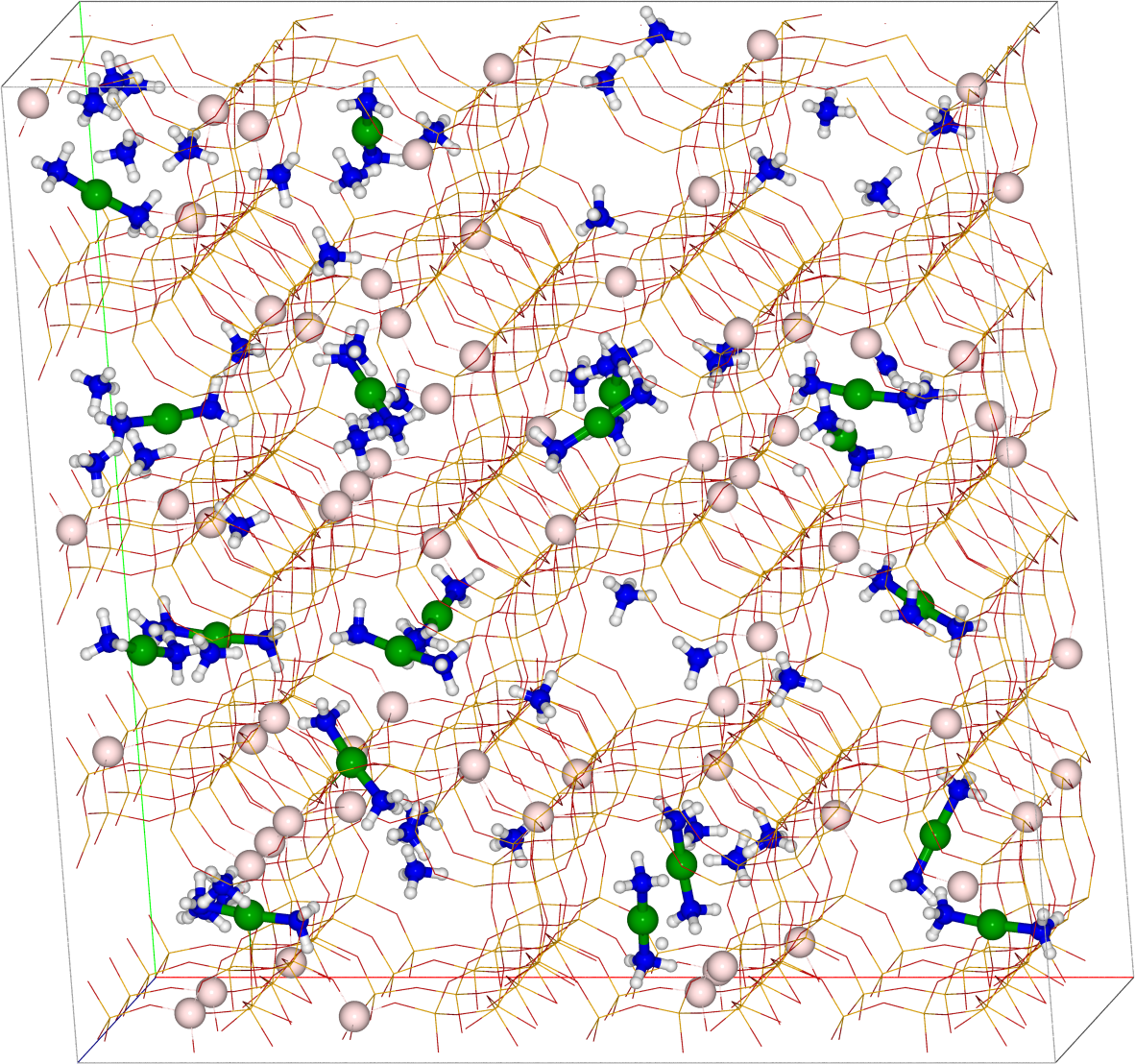}
    \caption{HR}
\end{subfigure}
\begin{subfigure}[b]{.24\textwidth}
\centering
    \includegraphics[width=\textwidth]{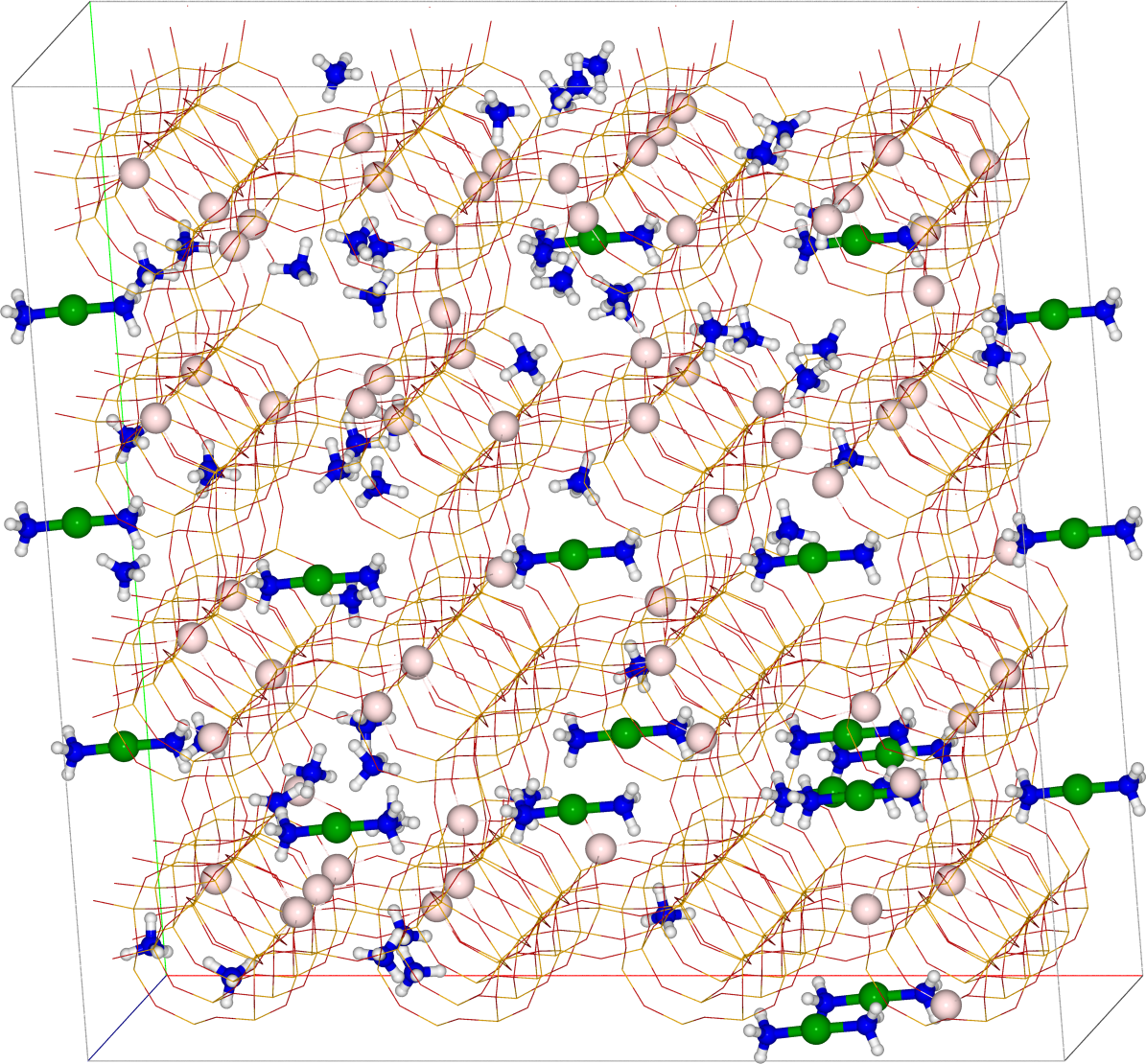}
    \caption{HB}
\end{subfigure}
\caption{\textbf{Structural  models used in NNP unbiased MD simulations. }Snapshots of the input structures corresponding to the eight different zeolite framework compositions considered. The first letter of the label indicates the Al content, L (low) Si/Al=28.5, M (medium) Si/Al=14.3 and H (high) Si/Al=10.3. The number in the top models indicates the Cu loading, 4 or 20 atoms per unit cell. The second letter in the bottom models that  contain 20 Cu atoms per unit cell indicates the Al distribution, forming pairs in 6R or 8R,  random (R) and biased (B). Si, O, Al, H, Cu and N atoms are depicted as orange, red, light brown, white, green and blue.}
\label{fig:snapshots_222_supercell_2600_atoms}
\end{figure}

A large \ce{T768O1536} supercell of dimension $\sim$37 \text{\AA} was used to construct eight models representing three Al contents, low (L), medium (M) and high (H),  two copper loadings (4 and 20 atoms per simulation box), and varying \ce{NH3}, \ce{NH4+} and \ce{H+} concentrations and Al pairings as described in Methods Section, Fig \ref{fig:snapshots_222_supercell_2600_atoms} and Table \ref{tab:dataset_composition_MD}. Unbiased MD simulations at 500 K were conducted for at least 5 ns on each ~2000-atom system.

\begin{figure}[!ht]
\captionsetup[subfigure]{labelformat=empty}
\centering
\begin{subfigure}{1.0\textwidth}
\centering
    \includegraphics[width=\textwidth]{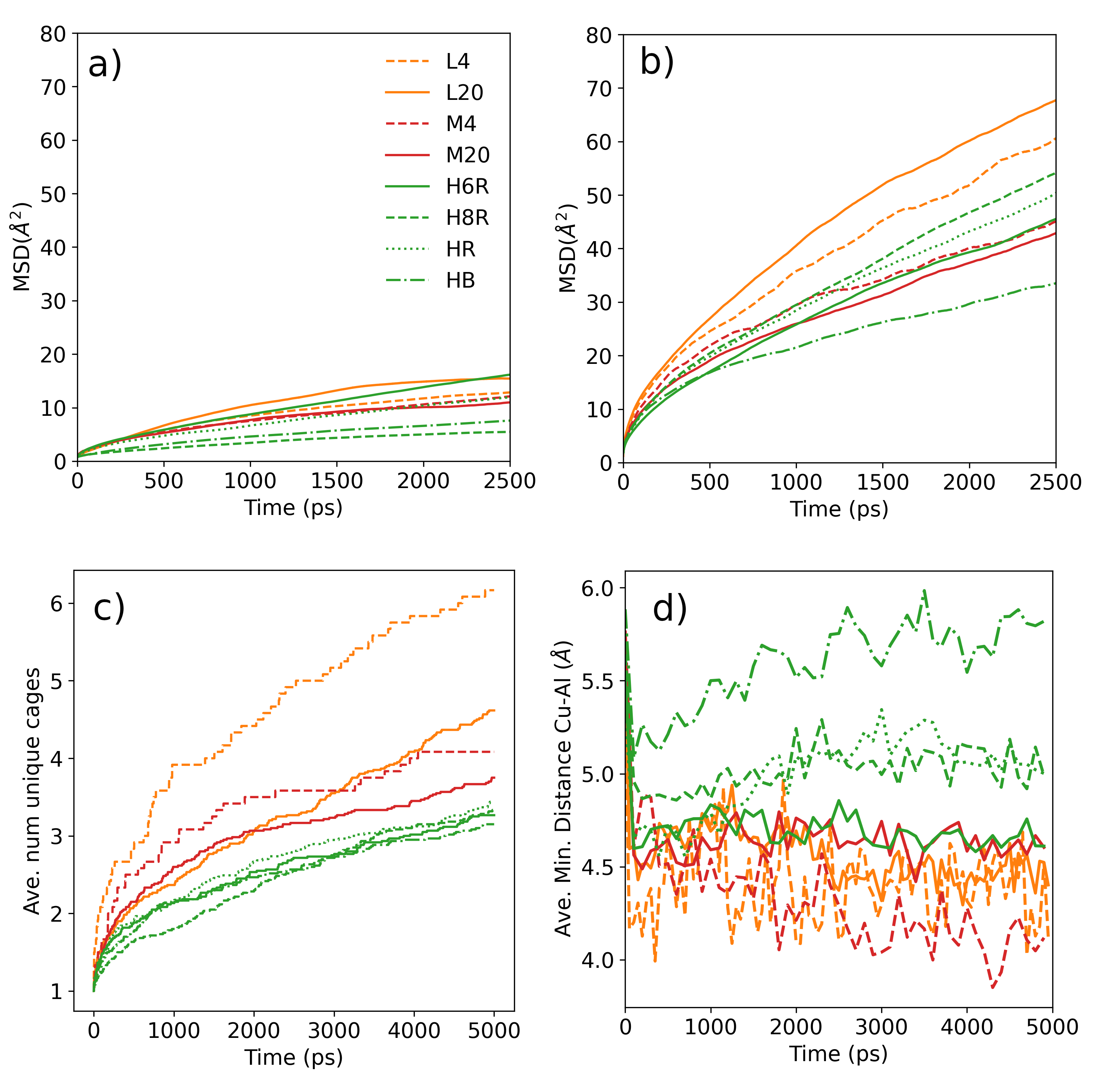}
\end{subfigure}%

\caption{\textbf{Mobility of} \ce{NH4+}  \textbf{and} \ce{[Cu(NH3)2]+}. (a) Mean square displacement (MSD) of N atoms in \ce{NH4+} (b) MSD of Cu atoms in the \ce{[Cu(NH3)2]+} complexes obtained from NPP-based unbiased MD simulations at 500 K. (c) Number of distinct cages visited by the Cu atoms. The apparent contradiction with the MSD plots is explained by the fact that the diffusion of \ce{[Cu(NH3)2]+} does not occur along a preferential direction. One Al pair in a 8R can accelerate the diffusion of \ce{[Cu(NH3)2]+} between two neighboring cages in multiple forward and backward steps, so that only two distinct cages are visited by this complex. (d) average minimum Cu-Al distances. All plots were obtained from the same simulations. MSD was calculated using eq \ref{eq:msd}.}
\label{fig:allmsd_and_cages_visited}
\end{figure}

Figures \ref{fig:allmsd_and_cages_visited} a and b track, respectively, the time evolution of the mean square displacements (MSDs) of the N atoms in the \ce{NH4+} cations and of the Cu atoms in the \ce{[Cu(NH3)2]+} complexes.  While both species have a net charge of +1, \ce{[Cu(NH3)2]+} is much more mobile while \ce{NH4+} cations are more closely attached to the \ce{AlO4-} units (Figs \ref{fig:scatterplots_Cu} and \ref{fig:scatterplots_nh4}). This is due to the positive charge in \ce{[Cu(NH3)2]+} complexes being highly shielded by the two \ce{NH3} ligands. 

The similarity among the MSD profiles of N atoms suggests that the mobility of \ce{NH4+} cations is rather independent of zeolite framework composition and Cu content  (Figure \ref{fig:allmsd_and_cages_visited}a), while the MSD traces for Cu (Figure \ref{fig:allmsd_and_cages_visited}b) suggest slightly lower mobility of \ce{[Cu(NH3)2]+} in the systems with high Al content (Si/Al $\sim$ 10, Figure \ref{fig:allmsd_and_cages_visited}b). The trends are similar in the number of distinct cages visited by the \ce{[Cu(NH3)2]+} complex over time (Figure \ref{fig:allmsd_and_cages_visited}c), but they are more stratified and more clearly show an increase in mobility with decreasing copper content (orange and red dashed lines higher than solid in Fig \ref{fig:allmsd_and_cages_visited}c). In 5 ns, each \ce{[Cu(NH3)2]+} complex visits on average fewer than three different cages in the models with high Al content, which increases to 3-4 for intermediate Al, and reaches a maximum of over 5 different cages visited for the L4 model, which has the lowest Al content and thus few Al pairs in 8R. This is in apparent contradiction to the biased simulations results for small models, that showed lower hopping free energy barrier for Al pairing in 8R. A potential explanation is that the hopping landscape is statically and dynamically heterogeneous, with the local chemical environment of each initial and final cages, and transient cage occupation by other mobile molecules influencing the mobility of copper complexes. Fig \ref{fig:scatterplots_Cu} shows the diversity in length and tortuosity in example trajectories of Cu atoms inside the zeolite microporous structure. While MSD profiles representing the average movement of the \ce{[Cu(NH3)2]+} species are relatively similar across catalyst models ($\sim$40-60 \text{\AA}$^2$ ), the local mobility of each individual \ce{[Cu(NH3)2]+} complex depends on its local chemical environment.

Since all framework \ce{AlO4-} units are compensated with either \ce{[Cu(NH3)2]+} or \ce{NH4+}, lower Cu content implies larger \ce{NH4+} concentration for a given Si/Al ratio. Comparison of L4 with L20, and M4 with M20 in Figure \ref{fig:allmsd_and_cages_visited}c suggests that \ce{NH4+} assists the long-range diffusion of \ce{[Cu(NH3)2]+}. A proposed explanation is that adsorbed \ce{NH4+} shields the attractive interaction between \ce{AlO4-} anionic sites  and the \ce{[Cu(NH3)2]+} complexes.  Each \ce{NH4+} forms two strong hydrogen bonds with the \ce{AlO4-} site, hence the lower mobility of \ce{NH4+} while \ce{[Cu(NH3)2]+} interacts with the zeolite through the H atoms of the coordinating \ce{NH3} molecules. This "crowding" of the anionic sites by the harder \ce{NH4+} can be observed statistically in the simulations. The average distances between the Cu atoms and the closest framework Al atoms plotted in Figure \ref{fig:allmsd_and_cages_visited}d are $\sim$ 4.5 \text{\AA} in the L and M models (orange and red lines), and increase to $\sim$ 5.0 \text{\AA} in the systems with higher Al content and thus a larger amount of charge-balancing \ce{NH4+} (green lines). The HB model has a heterogeneous Al distribution and its anomalously high Cu-Al distance is due to \ce{[Cu(NH3)2]+} complexes in Al-poor regions.  
 
Previous studies have suggested that \ce{[Cu(NH3)2]+} migration is fast only between the three cages sharing a common framework Al, while long-range diffusion to non-adjacent cages is limited to $\sim$ 9 \text{\AA} due to decaying electrostatic to the \ce{[Cu(NH3)2]+} and the anionic \ce{AlO4-} site. \cite{Paolucci2017a, Millan2021,Krishna2023InfluenceNH3}. This argument is rigorously true when the final cages contain no additional Al sites, as our model systems from the first section. In real systems, however, the long-range diffusion is easily explained through a sequence of local steps combining the crossing of 8R windows into adjacent Al-containing cages, followed by the exchange of \ce{NH4+} as compensating cation (Scheme \ref{fig:scheme_migration} in the Supporting Information). The low mobility of \ce{NH4+} revealed by the present simulations suggests that a limiting factor for such long-range diffusion of \ce{[Cu(NH3)2]+} complexes is the slower rate of counter-migration of the charge-compensating \ce{NH4+}. While \ce{[Cu(NH3)2]+} could act as a migrating compensating cation, this results in no net migration of copper.  

To explore this hypothesis, two additional MD simulations of 5 ns were run using two modified M20 models, one of them containing 30 protons as compensating cations, labeled M20-H+, and another one with 60 additional \ce{NH3} molecules added to the system, labeled M20-NH3. Protons on the Brønsted acid sites are fairly static and localized within the four O atoms directly attached to Al, so long-range charge-compensation diffusion of \ce{H+} should be hindered.  In contrast, additional \ce{NH3} should facilitate the  movement of the positive charges via proton transfer from \ce{NH4+} to \ce{NH3} via a Grotthuss-like chain of proton transfers, thus allowing faster charge compensation between separated \ce{AlO4-} units without the need of physically displacing the strongly attached \ce{NH4+} cations. 

\begin{figure}[H]\captionsetup[subfigure]{labelformat=parens, justification=centering}
\centering
\begin{subfigure}{1.0\textwidth}
\centering
    \includegraphics[width=\textwidth]{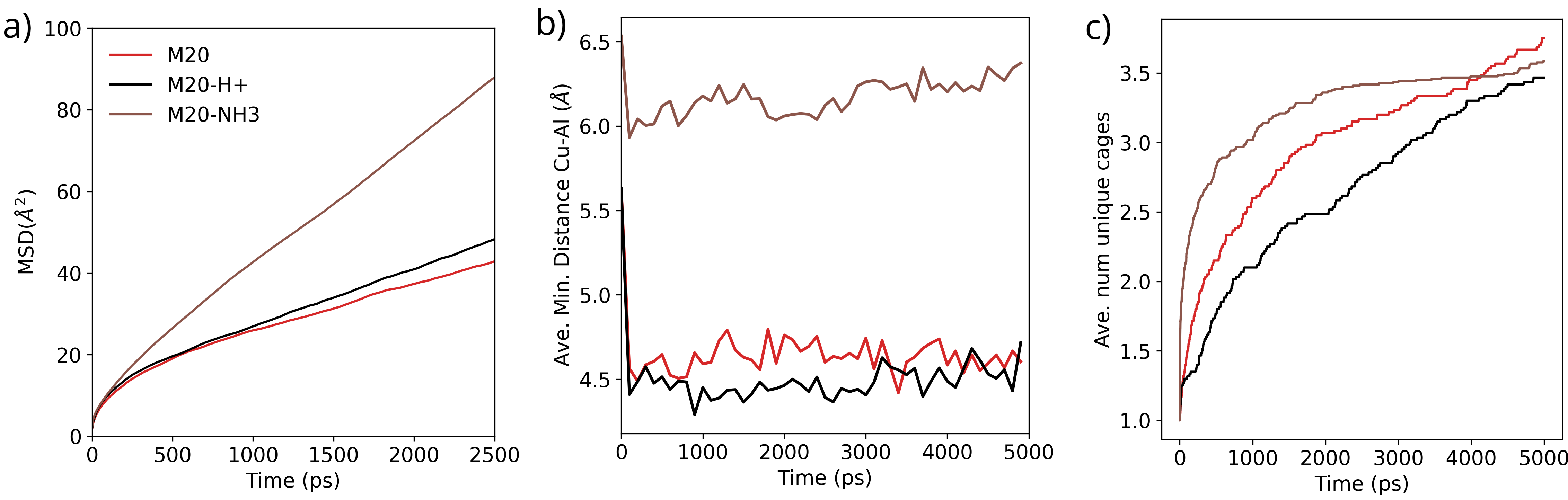}
\end{subfigure}%
\caption{\textbf{Influence of framework charge shielding on the mobility of } \ce{[Cu(NH3)2]+}. (a) Mean square displacement (MSD) of Cu atoms in the  \ce{[Cu(NH3)2]+} complexes, (b) average minimum Cu-Al distances and (c) number of distinct cages visited obtained from NPP-based unbiased MD simulations at 500 K in models with different charge-compensating cations and additional \ce{NH3} molecules. MSDs where calculated using eq \ref{eq:msd}.}
\label{fig:msds_Hs_NH3_NH4}
\end{figure}

 The plots in Figure \ref{fig:msds_Hs_NH3_NH4} confirm this hypothesis. The MSD of Cu atoms in Figure \ref{fig:msds_Hs_NH3_NH4}a does not change when \ce{NH4+} cations are substituted by protons (black lines) with a similar electrostatic shielding effect as \ce{NH4+}. However, copper mobility increases significantly in the presence of excess \ce{NH3} molecules (brown lines), suggesting that free \ce{NH3} facilitates the charge re-equilibrium following \ce{[Cu(NH3)2]+} crossing 8R. In keeping, the average  distances between the Cu atoms and the closest framework Al atoms increases from  $\sim$ 4.5 \text{\AA} to $\sim$ 6.0 \text{\AA} with the added \ce{NH3}  (Figure \ref{fig:msds_Hs_NH3_NH4}b). 
Lastly, the number of distinct cages visited by \ce{[Cu(NH3)2]+} (Figure \ref{fig:msds_Hs_NH3_NH4}c) indicates again a slightly lower long-range mobility of \ce{[Cu(NH3)2]+} in the model containing Brønsted acid sites and enhanced diffusion in the presence of an excess of \ce{NH3}.

Our results provide theoretical backing to recent experimental observations of the enhancing effect of gas phase \ce{NH3} during the solid-state ion-exchange of copper using mixtures of copper oxides and zeolites, which allows the fast preparation of Cu-exchanged zeolites at low temperatures (473-523 K) \cite{Shwan2015Solid-StateTemperature, Vennestrm2019TheZeolites}. After $\sim$ 4000 ps the three models reached similar steady states with $\sim$ 3.5 distinct cages being visited by each complex, corresponding to the final state of the ion-exchange process with a random distribution of \ce{[Cu(NH3)2]+} complexes occupying the whole unit cell.

\subsection{Bimolecular complexes and mechanistic implications for the \ce{NH3}-SCR-NOx reaction}
According to the proposed mechanism,\cite{Paolucci2017a} the reaction rate depends directly on the probability of finding simultaneously two \ce{[Cu(NH3)2]+} complexes in the same cage, which we analyzed separately from the mobility calculations. 
 
\begin{figure}[H]
\captionsetup[subfigure]{labelformat=parens, justification=centering}
\centering
\begin{subfigure}{\textwidth}
\centering
    \includegraphics[width=\textwidth]{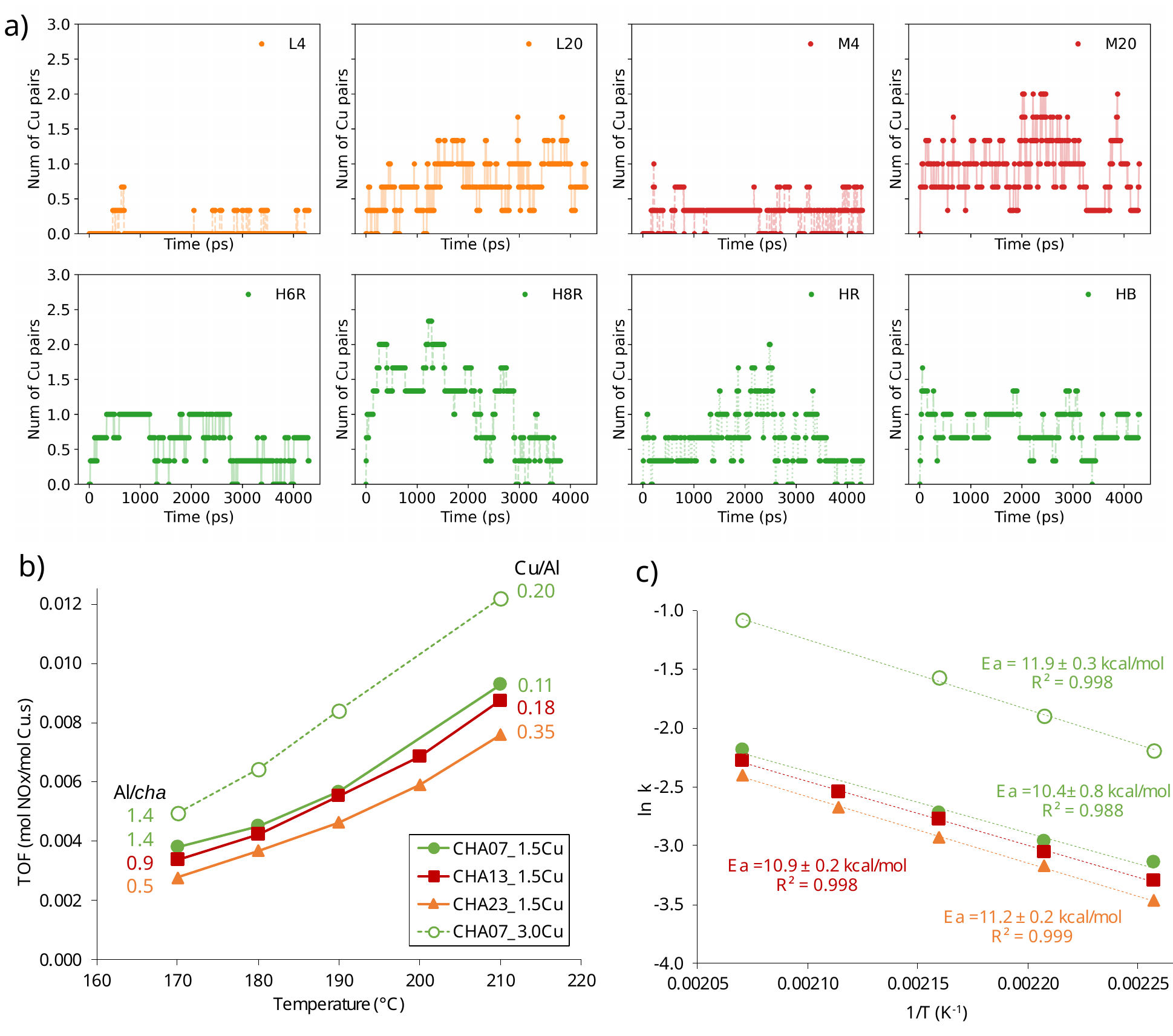}
\end{subfigure}%
\caption{\textbf{Probability of formation of the active sites and catalytic results for the \ce{NH3}-SCR-NOx  reaction}. (a) Time evolution of the average number of \ce{[Cu(NH3)2]+} pairs in the same cage for the eight systems in Figure \ref{fig:snapshots_222_supercell_2600_atoms}. The geometric center of every cage in the unit cell are determined and two Cu atoms are considered to be in the same cage when both have the same cage center as the nearest cage center. (b) Turnover frequency (TOF) values on a per Cu-ion basis for the $\sim$1.5\%wt Cu-containing zeolites with different Si/Al molar ratios (b) and CHA07 (Si/Al$\sim$7) zeolite with two different Cu-contents ($\sim$1.5 and $\sim$3 \%wt Cu) (c).}
\label{fig:cu_pairs_and_tof}
\end{figure}

Figure \ref{fig:cu_pairs_and_tof}a tracks the time evolution of the average number of \ce{[Cu(NH3)2]+} pairs present simultaneously within the same \textit{cha} cage. As expected, the probability of having two \ce{[Cu(NH3)2]+} complexes in the same cage directly correlates with the total amount of Cu in the system (compare L4 with L20 or M4 with M20 models). More interestingly, for a given Cu loading, the average number of \ce{[Cu(NH3)2]+} pairs in the same cage increases with increasing the Al content in the system (compare L4 with M4 or L20 with M20). This trend runs opposite to our observation about the number of distinct cages visited, which was highest for the model with lowest Al and lowest Cu content. That is, the chemical environments that result in a higher rate of long-range diffusion of copper complexes do not result in formation of more pairs, or by extension in higher reactivity, . 
 
As regards the role of Al distribution, the  four models with a Si/Al ratio $\sim$10 and different Al pairings (H6R, H8R, HR, HB) (see Methods Section, Fig \ref{fig:snapshots_222_supercell_2600_atoms} and Table \ref{tab:dataset_composition_MD})  behave similarly and the plots in Figure \ref{fig:cu_pairs_and_tof}a only suggest a lower probability of finding two \ce{[Cu(NH3)2]+} in the same cage in the H6R model with Al pairs in the 6R windows, and perhaps a slightly higher probability of bringing two \ce{[Cu(NH3)2]+} complexes together in the H8R model with Al pairs in the 8R windows. The increased \ce{NH3} loading in M20-NH3 also results in increased number of pairs (Figure \ref{fig:cu_pairs_nh3}). 

\subsection{Experimental validation for the low-T \ce{NH3}-SCR-NOx  reaction catalyzed by Cu-CHA zeolites with controlled composition }

To experimentally validate the computational predictions, three CHA samples with different Si/Al molar ratios ranging from 7.3 to 23.3, which translates to a broad range of 1.4 to 0.5 Al sites per \textit{cha} (see Table \ref{tab:properties_synthesize_cha} in the Supporting Information) were synthesized as described in Methods. Then, the same Cu loading ($\sim$1.5\%wt Cu) was introduced within the three CHA materials, which resulted in a similar amount of initial Cu atoms per \textit{cha} cage, $\sim$0.17, but different Cu/Al ratios (from 0.11 to 0.35, see Table \ref{tab:chemcomp_cu_exchange_cha}). In addition, the CHA sample with Si/Al ratio $\sim$7.3 was also loaded with 3.0\%wt Cu resulting in an additional sample with increased amount of initial Cu atoms per \textit{cha} cage (0.3). 

The catalytic tests to evaluate the low-temperature SCR-NOx activity of the different Cu-CHA materials were performed at very high space velocities (1800000 ml/h.grcat)  to assure low NO conversions ($<$20\%). Turnover frequency values (TOF) were obtained for each Cu-CHA sample at five different temperatures (see Figure \ref{fig:cu_pairs_and_tof}b). The TOF values obtained for the three catalysts with different Si/Al molar ratio and the same Cu content ($\sim$1.5\%wt Cu) exhibit a continuous activity enhancement as the Al/\textit{cha} cage ratio increases from 0.5 to 1.4 (see Figure \ref{fig:cu_pairs_and_tof}b), in agreement with the theoretical conclusion that the probability of finding simultaneously two \ce{[Cu(NH3)2]+} complexes in the same \textit{cha} cage increases with increasing the Al content in the zeolite system. Comparison of the TOF values obtained for samples with the same Si/Al ratio and different Cu content (full and dotted green lines in Figure \ref{fig:cu_pairs_and_tof}b) or even with similar Cu/Al ratio and different Cu content (red and dotted green line in Figure \ref{fig:cu_pairs_and_tof}b) confirm that the catalytic activity clearly improves with increasing the Cu/\textit{cha} ratio, in good agreement with the theoretical conclusion that the probability of forming \ce{[Cu(NH3)2]+} pairs in the same \textit{cha} cage directly correlates with the total amount of Cu in the system.  


The experimental apparent activation energies (Figure \ref{fig:cu_pairs_and_tof}c) are similar for all samples irrespective of the Si/Al ratio, Cu content or catalytic activity, ranging from 10.4 $\pm$ 0.8 to 11.9 $\pm$ 0.3 kcal/mol. The catalytic activity measured by the TOF and normalized by Cu content, however, increases in parallel with the calculated likelihood of two-copper encounters in the same cage (Fig \ref{fig:cu_pairs_and_tof}a). This supports the argument that the formation of \ce{[Cu(NH3)2]+} pairs in the same \textit{cha} cage is responsible for the generation of the binuclear active sites that catalyze the reaction, although copper diffusion may not necessarily be the rate-determining step of the global process.

\section{Conclusions}

Biased and unbiased MD simulations using a newly trained NNP have achieved high accuracy, chemical diversity, length- and time scale, allowing the systematic investigation of the influence of catalyst composition and adsorbed \ce{NH3} on the mobility of \ce{[Cu(NH3)2]+}  cations in Cu-CHA catalysts. 

Biased simulations on small systems showed that single \ce{[Cu(NH3)2]+} cation hops between adjacent \textit{cha} cages are very sensitive to the Al distribution and, in general, Al pairs in 8R windows lower the free energy barrier for diffusion and stabilize the product configuration with two \ce{[Cu(NH3)2]+} cations in the same cage. This might be taken to suggest that the rate of the SCR-NOx reaction could be accelerated by selectively positioning the Al atoms as Al pairs. However, even though those results are well beyond the limits of traditional AIMD, the simulation cells employed are overly simple and lack realistic Al and \ce{NH4+} concentration. 

Unbiased MD simulations using time scales of multiple nanosecods and supercells with over 2300 framework atoms at a variety of Si/Al ratios, \ce{Cu+}, \ce{NH4+} and \ce{NH3} loadings show that \ce{[Cu(NH3)2]+} cations can visit on average 3-4 cages and diffuse as far as 30 \text{\AA} in a few nanoseconds. They also show that the long-range migration to remote cages requires the simultaneous displacement of a charge-compensating \ce{NH4+} cation. An excess of \ce{NH3} facilitates the movement of the positive charges via proton transfer from \ce{NH4+} to \ce{NH3}, thus enhancing the long-range diffusion of \ce{[Cu(NH3)2]+} complexes.

Regarding catalytic activity, we observed that the probability of finding two \ce{[Cu(NH3)2]+} complexes in the same cage, which is necessary for the SCR-NOx reaction at low temperature, correlates directly with the Cu content and on the Al content but not so much on the Al distribution. These trends were confirmed experimentally through testing the SCR-NOx reaction at low temperatures using Cu-CHA zeolites with different Si/Al and Cu/Al molar ratios, where we found increasing catalytic performance with increasing Al and Cu loading.

These results demonstrate the power of combining high-throughput DFT calculations, machine learning and molecular dynamics simulations for simulating transport in nanoporous catalysts. The collection of the training data and the training of the NNP had a lower total computational cost than a single traditional AIMD simulation and resulted in scalable, fast and accurate simulations. 

This overall strategy is broadly applicable to other unsolved questions in nanoporous catalysts, since it enables DFT-accuracy, nanosecond-long MD simulations for thousands of atoms, and possibly beyond, by combining ML and enhanced sampling techniques\cite{Sipka2023DifferentiableEvents}.

\section{Methods}
\subsection{Catalyst Models}
The Cu-CHA catalytic system was modeled using three different supercells of increasing size (see Figure \ref{fig:snapshots_cell_sizes}).
The smallest system is a 1x1x1 hexagonal supercell of CHA framework containing 2Al, 34 Si and 72 O atoms, with lattice parameters, a = b = 13.81 \text{\AA}, c = 15.00 \text{\AA}, $\alpha$ = 89.86\degree, $\beta$ = 89.94\degree, $\gamma$ = 120.41\degree. 
Three models with different lattice distribution of the two Al atoms were created, and the two negative charges introduced by the presence of Al in the framework were compensated with two \ce{[Cu(NH3)2]+} complexes (Figure \ref{fig:snapshots_of_alpairs_and_nn_vs_dft}a). 
These models were used to perform DFT-based umbrella sampling (US) molecular dynamics (MD) simulations and in the first generation of the NNP.

A 2x2x2 triclinic supercell containing 96 T and 192 O atoms, with lattice parameters a = 18.68 \text{\AA}, b = 18.67 \text{\AA}, c = 18.67 \text{\AA}, $\alpha$ = 94.68\degree, $\beta$ = 94.63\degree, $\gamma$ = 94.72\degree  was used for the subsequent training of the NN potentials including additional Al distributions, Al contents, and compensating species (\ce{[Cu(NH3)2]+}, \ce{NH4+} and \ce{H+}) as well as additional \ce{NH3} molecules as summarized in Table \ref{tab:dataset_composition_training}.
For the systems with 3 and 7 Al atoms, ten random Al distributions were generated. For the systems with two Al atoms ten different Al distributions were generated, one with 2 Al in the same 4R, one with two Al in the same 6R, four of them containing 2 Al in the same 8R, and four of them with the 2 Al in different rings. Single-point DFT calculations and NNP US-MD simulations were performed using this model.  

Finally, a 4x4x4 triclinic supercell containing 768 T and 1536 O atoms, with lattice parameters a = 37.35 \text{\AA}, b = 37.38 \text{\AA}, c = 37.34 \text{\AA}, $\alpha$ = 94.64\degree, $\beta$ = 94.59\degree, $\gamma$ = 94.47 was employed to run NNP unbiased MD simulations on more realistic systems and reaction conditions. Three lattice compositions corresponding to Si/Al $\sim$ 30 (26 Al, 742 Si and 1536 O atoms), Si/Al $\sim$ 15 (50 Al, 718 Si and 1536 O atoms) and Si/Al $\sim$ 10 (50 Al, 718 Si and 1536 O atoms) were considered, and for each of them three random Al distributions were generated. For the model with the highest Al content (Si/Al $\sim$ 10), three additional Al distributions were generated in which all Al atoms were either forming pairs in the 6R units, forming pairs in the 8R windows, or heterogeneously distributed along the system so that Al-rich and Al-poor regions are found in the same unit cell. The negative charges in each model were compensated with \ce{[Cu(NH3)2]+}, \ce{NH4+} and \ce{H+} as summarized in Table \ref{tab:dataset_composition_MD} in the Supporting Information. 
In the models with low and medium Al content, two Cu loadings were considered, 4 and 20 atoms per unit cell, corresponding to Cu spatial densities of 0.08 and 0.4 Cu/1000 \text{\AA}$^2$ respectively and 0.5\% and 2.6\% Cu/Si ratios (systems labeled L4, L20, M4 and M20, Figure \ref{fig:snapshots_222_supercell_2600_atoms}). In the models with high Al content, the Cu loading was always 20 atoms per unit cell, and the labels in Figure \ref{fig:snapshots_222_supercell_2600_atoms} indicate the four different Al distributions considered, with Al pairs in the same 6R (H6R), with Al pairs in the same 4R (H4R), random distribution (HR) and a biased, spatially-heterogeneous distribution containing regions with high and low Al content (HB). 

All DFT calculations (single points evaluations as well as the molecular dynamics simulations) were carried out at the revPBE+D3 \cite{Zhang1998,Grimme2010} level of theory with the software CP2K \cite{Vandevondele2005Quickstep:Approach}. The Gaussian and Plane Waves (GPW) method \cite{Lippert1999TheSimulations} was used with the TZVP basis set for all atoms except Cu, which was described with the DZVP-MOLOPT-SR-GTH basis set. A cutoff energy of 400 Ry was used for the auxiliary plane waves and the core electrons were represented with GTH pseudopotentials \cite{Goedecker1996SeparablePseudopotentials}.

All NNP-MD simulations were performed using NNPs trained with the PaiNN\cite{Schutt2021} architecture. Regular NNP-MD simulations were run in the NVT ensemble at 500 K, and consisted of a production run of at least 3 ns after 100 ps of equilibration. The temperature was controlled by a Nosé–Hoover chain thermostat \cite{Martyna1992, Hoover1982} with three beads and a time constant of 200 fs. The time step to integrate the equations of motion was set to 0.5 fs.

The mean squared displacements (MSD) obtained for \ce{[Cu(NH3)2]+} and \ce{NH4+} were calculated via the following equation:
\begin{equation}
  MSD(t) = \frac{1}{t_s - t + 1} \sum_{t_0=0}^{t_s=t} \frac{1}{N} \sum_{i=1}^{N} \left [ r_i(t_0+t) - r_i(t_0)\right ]^2 
  \label{eq:msd}
 \end{equation}
where N corresponds to the number of \ce{[Cu(NH3)2]+} or \ce{NH4+} cations, $t_s$ is the number of time steps in the simulation, the $t_0$ are the different time origins.


Gibbs free energy profiles for the diffusion of \ce{[Cu(NH3)2]+} species through the 8R windows of Cu-SSZ-13 zeolite were obtained from umbrella sampling (US) simulations \cite{Torrie1977} at 423 K. The success of the US technique relies on the choice of a collective variable (CV) that represents unambiguously the process considered. It has been shown in the literature that the collective variable ($\xi$) that best describes the diffusion of \ce{[Cu(NH3)2]+} from one cavity A ($\xi < 0$ ) to another one B ($\xi > 0$ ) is the projection of the position vector of the Cu atom on the vector normal to the average plane of the 8R (see Figure \ref{fig:cv}). This collective variable takes the value $\xi = 0$  when the Cu atom is located in the plane of the 8R. In the biased simulations, the collective variable $\xi$  was split into 40 (for DFT-based simulations) or 80 (for NN-based simulations) equidistant windows in the range $\xi$ = -4 to $\xi$ = 4. In each window, an independent biased MD simulation of 20 ps was run restricting the collective variable to one specific value but ensuring complete sampling of the configuration space in all other degrees of freedom. A harmonic bias potential centered at the corresponding value of the collective variable was applied to restrict the sampling to each window individually and to ensure sufficient overlap between the sampling of adjacent windows. The force constant of the harmonic potential was set to 12 kcal/mol/\text{\AA}$^2$. All DFT-based biased simulations were performed with PLUMED\cite{Kumar1992} interfaced to the CP2K engine.

\subsection{Neural network potential training}
All NN potentials trained were based on the PaiNN architecture \cite{Schutt2021} which uses equivariant message-passing for the ground truth prediction. For the training we used three convolutions, a cutoff radius of 5\text{\AA}, 32 Gaussian functions for distance expansion and 128-size vector for the atomic features. Neural network parameters were optimized using Adam algorithm with a batch size of 32, initial learning rate of \num{1e-4}, learning decay of 0.5 and learning patience of 25. 

The reference energies of Si, Al, Cu, N, O, and H were evaluated using the revPBE+D3 energies of the lowest energy structure of each composition. These energies were used to fit a linear regression model with atomic contribution to the reference energy of each element as predictor variables. The reference energy was subtracted from the revPBE+D3 energy of every geometry in the training, test and validation sets. The forces were used without modifications. The training coefficients for the energies and the forces were set to 0.01 and 1.0, respectively.

The acquisition of training data was performed using active learning (AL) with a query-by-committee approach\cite{Behler2015, smith2018less, Schran2020, Musil2019, Peterson2017b, Lookman2019, Shapeev2020a, Imbalzano2021} as illustrated in Figure \ref{fig:fig1}a. In this approach a committee (ensemble) of NNPs is trained on the available labelled data and new data is collected based on the maximum disagreement (variance) of the prediction the committee members. 

The first generation of the potential was trained on a randomly collected subset of the DFT data generated from a previous study \cite{Millan2021} and from three biased simulations performed with DFT at 423 K, used as reference ground truth. In total, there were $\sim$9000 geometries in the initial dataset. This pretrained potential was then retrained in 4 active learning loops using the 2x2x2 triclinic supercell described in previous section (see Figure \ref{fig:snapshots_cell_sizes}. For each loop, biased MD trajectories were generated with the learned interatomic potential of the previous loop.  The following temperatures were used in the active learning loops 298 K, 423 K, 500 K and 550 K.  The bias was applied using a harmonic potential on the collective variable used in US. Each MD was run for 100 ps and geometries were collected every 100 fs. The selection of the new geometries from the MD trajectories was carried out using as criterion the force uncertainty from an ensemble of three NN potentials. The variances of the forces from the ensemble of potentials were ranked in descending order and the first geometries were selected to increase the dataset in $\sim$10\% . The nonphysical geometries and those with low uncertainty, $<$ 2 kcal/mol were discarded.
Up to this point, the dataset contained structures where the Al substitutions where compensated with \ce{Cu+} as diaminecopper(I) so that the trained potential did not properly describe local environments of Al compensated with \ce{NH4+} or \ce{H+}. The acquisition of new geometries with new compositions including \ce{NH4+} and \ce{H+} was performed using adversarial attack \cite{Schwalbe-Koda2021} for 6 more generations with NNP trained on the last generation of AL.
The initialization of adversarial attacks were performed by displacing each coordinate $\sigma$ $\sim$N(0,0.01 \text{\AA}) of one optimized geometry for every chemical composition containing \ce{NH4+} and \ce{H+}. 
The resulting attack $\sigma$ was optimized for 100 iterations using the Adam optimizer with learning rate of 0.0001 and with the normalized temperature kT set to 20 kcal/mol. 

Then, five more generations of active learning was performed, with biased MD simulations at temperatures ranging from 600 to 1000 K. All models were trained by randomly partitioning the available data into training (60\%), validation (20\%) and test sets (20\%). This splitting was perform for each composition and then the training, test and validation sets of each composition were combined in the final training, test and validation sets. Thus, the final training, test and validation sets had a balanced distribution of all chemical compositions.

\subsection{Zeolite synthesis} \label{synthesis}
\subsubsection{CHA zeolite with a Si/Al$\sim$7 (CHA07) }
5.91 g of a 25 wt\% aqueous solution of N,N,N-trimethyl-1-adamantylammoniumhydroxide (TMAda, Sachem) was added to 5.41 g of deionized \ce{H2O}. Next, 0.11 g of \ce{Al(OH)3} (76.5 wt\%, Thermo Fisher) and 0.85 g of a 5M sodium hydroxide solution (NaOH: 16.7 wt\% NaOH in deionized water; NaOH pellets 98 wt\%, Alfa Aesar) was added to the aqueous TMAdaOH solution and the mixture was stirred under ambient conditions for 15 minutes. Finally, 2.13 g of colloidal silica (Ludox HS40, 40 wt\%, Sigma Aldrich) was added to the mixture and stirred for 2 h under ambient conditions. The final gel composition was: \ce{SiO2} : 0.056 \ce{Al2O3} : 0.49 TMAdaOH : 0.25 NaOH : 46.5 \ce{H2O}. The resultant gel was charged into a stainless steel autoclave with a Teflon liner. The crystallization was then conducted at 160 \textdegree C for 6 days under dynamic conditions. The solid product was filtered, washed with abundant water, and dried at 100 \textdegree C. The solids were calcined at 580 \textdegree C for 5 h in air. 

\subsubsection{CHA zeolite with a Si/Al$\sim$13 (CHA13) }
5.89 g of a 25 wt\% aqueous solution of N,N,N-trimethyl-1-adamantylammoniumhydroxide (TMAda, Sachem) was added to 5.42 g of deionized \ce{H2O}. Next, 0.042 g of \ce{Al(OH)3} (76.5 wt\%, Thermo Fisher) and 0.86 g of a 5M sodium hydroxide solution (NaOH: 16.7 wt\% NaOH in deionized water; NaOH pellets 98 wt\%, Alfa Aesar) was added to the aqueous TMAdaOH solution and the mixture was stirred under ambient conditions for 15 minutes. Finally, 2.10 g of colloidal silica (Ludox HS40, 40 wt\%, Sigma Aldrich) was added to the mixture and stirred for 2 h under ambient conditions. The final gel composition was: \ce{SiO2} : 0.02 \ce{Al2O3} : 0.50 TMAdaOH : 0.26 NaOH : 47\ce{H2O}. The resultant gel was charged into a stainless steel autoclave with a Teflon liner. The crystallization was then conducted at 160 \textdegree C for 6 days under dynamic conditions. The solid product was filtered, washed with abundant water, and dried at 100 \textdegree C. The solids were calcined at 580 \textdegree C for 5 h in air.

\subsubsection{CHA zeolite with a Si/Al$\sim$23 (CHA23) }
1.55 g of FAU zeolite (FAU, CBV760 with Si/Al=26, Zeolyst, lot number: 76004N002648) was added to 6.74 g of a 25 wt\% aqueous solution of N,N,N-trimethyl-1-adamantylammoniumhydroxide (TMAda, Sachem). The mixture was maintained under stirring the required time to evaporate the excess of water until achieving the desired gel concentration. The final gel composition was: \ce{SiO2} : 0.019 \ce{Al2O3} : 0.4 TMadaOH : 5 \ce{H2O}. The resultant gel was charged into a stainless steel autoclave with a Teflon liner. The crystallization was then conducted at 175 \textdegree C for 12 days under static conditions. The solid product was filtered, washed with abundant water, and dried at 100 \textdegree C. The solids were calcined at 580 \textdegree C for 5 h in air.

\subsubsection{Cu-exchange treatments }
The Na-containing CHA calcined solids were first exchanged with a 2M aqueous solution of ammonium nitrate (\ce{NH4Cl}, Sigma-Aldrich, 99\% by weight) with a liquid/solid ratio of 10, maintaining the mixture at 80 \textdegree C for 2 hours under agitation. Afterwards, the solids were recovered by filtration.
0.3 g of the Na-free zeolites was introduced in 30 ml of an aqueous solution of \ce{Cu(CH3COO)2}·\ce{H2O} [14.1 mg of \ce{Cu(CH3COO)2}·\ce{H2O} dissolved in 30 ml of water or 28.3 mg of \ce{Cu(CH3COO)2}·\ce{H2O} dissolved in 30 ml of water for materials containing $\sim$1.5 or 3\%wt Cu, respectively], maintaining a solid/liquid ratio of 10 g/l at 80 \textdegree C for 24 h. Finally, the solids were filtered and washed with distilled water, dried and calcined at 550 \textdegree C in air for 4 h.

\subsection{Characterization }
Powder X-ray diffraction (PXRD) measurements were performed with a multi sample Philips X’Pert diffractometer equipped with a graphite monochromator, operating at 40 kV and 35 mA, and using Cu K$\alpha$ radiation ($\lambda$ = 0.1542 nm). The PXRD patterns reveal the good crystallization of the CHA materials (see Figure  \ref{fig:pxrd_patterns} in the Supporting Information), all of them presenting particle sizes within the sub-micron scale (below 1 µm, see Figure  \ref{fig:pxrd_patterns} and Table \ref{tab:properties_synthesize_cha} in the Supporting Information).
Chemical analyses were carried out in a Varian 715-ES ICP-Optical Emission spectrometer, after solid dissolution in \ce{HNO3}/HCl/HF aqueous solution. 
Nitrogen adsorption isotherms at -196 \textdegree C were measured on a Micromeritics ASAP 2020 with a manometric adsorption analyser to determinate the textural properties of the samples.
The morphology of the samples was studied by field emission scanning electron microscopy (FESEM) using a ZEISS Ultra-55 microscope.

\subsection{Catalytic evaluation}
The catalytic activity was evaluated for the selective catalytic reduction (SCR) of NOx with \ce{NH3} in a fixed bed, quartz tubular reactor with $\sim$1 cm inner diameter. 20 mg sieve fractionated catalysts (200-400 {\textmu}m) were placed between 0.6 of silicon carbide fractions (SiC, Fisher Chemical, 200-400 {\textmu}m) generating an overall bed height of $\sim$ 1.2 cm. The catalysts were introduced in the reactor and heated up to 550 \textdegree C in a 300 mL/min flow of nitrogen and maintained at this temperature for one hour. Afterwards, the feed was admitted over the catalyst with an overall flow of 600 mL/min. The feed composition for the catalytic tests performed over the Cu-containing catalysts was 500 ppm NO, 550 ppm NH3, 7\% O2 and 5\% H2O, resulting in a very high space velocity (GHSV=1800000 ml/h.grcat). The evaluated reaction temperatures were 190, 180 and 170 \textdegree C. The conversion of NO was measured under steady state conversion at each temperature using a chemiluminescence detector (Thermo 62C). The TOF were estimated by dividing the moles of NO molecules converted per second by the moles of Cu atoms in the catalysts, according to

\begin{equation}
    TOF = \frac{moles\ NO\ converted/s}{moles\ Cu}
\end{equation}

From the low temperature NO conversion results, the rate constants (k) can be calculated using a first-order kinetic equation, as described previously in the literature: \cite{Long2002SelectiveStudy}

\begin{equation}
k = -\frac{F_0}{[NO]_0W}\ln(1-X)    
\end{equation}

where $F_0$ is the molar NO feed rate, $[NO]_0$ is the molar concentration at the inlet, W is the catalyst amount (gr) and X is the NO conversion.
The Arrhenius equation was employed to estimate the apparent activation energies (Ea) after its linearization as follows: 

\begin{equation}
k = Ae^{-E_a/RT} 
\end{equation}
\begin{equation}
   \ln{k} = \ln{A} - \frac{E_a}{R}\left(\frac{1}{T}\right) 
\end{equation}

where A is the pre-exponential factor, R is the universal gas constant and T denotes the absolute temperature associated with the reaction (in Kelvin).

\section{Data availability}
The experimental and computational data that support the findings of
this study are available from the corresponding author upon reasonable
request.
The datasets generated during this study are available at \url{https://figshare.com/projects/Dataset_and_machine_learning_potential_Cu-CHA/167645}. The code used for this study can be downloaded from https://github.com/learningmatter-mit/NeuralForceField.

\bibliography{references2}

\begin{acknowledgement}
R.M acknowledges the Margarita Salas grant from the Ministerio de Universidades, Spain, funded by the European Union-Next Generation EU. The authors are grateful for computation time allocated on the MIT SuperCloud cluster, the MIT Engaging cluster at the Massachusetts Green High Performance Computing Center (MGHPCC); and Summit at the Oakridge Leadership Computing Facility through the 2021 ALCC DOE program. R.G-B. thanks the Jeffrey Cheah Career Development Chair. R.M thanks Gavin Winter for assistance during the training of NNP and processing of the MD simulations, and Simon Axelrod for implementing PaiNN. M.B. and M.M. thank financial support from Spanish Government through CEX2021-001230-S, PID2020-112590GB-C21, PID2021-122755OB-I00 and TED2021-130739B-I00 (MCIN/AEI/FEDER, UE), and from CSIC through the I-link+ Program (LINKA20381). E.B. acknowledges the Spanish Government-MCIU for a FPI scholarship (PRE2019-088360). The Electron Microscopy Service of the UPV is acknowledged for its help in sample characterization.

\end{acknowledgement}

\section{Author Contributions}
R.G-B. and M. B. designed the project. R. M. developed the NNP and performed all simulations, supervised by R.G.-B. M. M. directed the zeolite synthesis and catalytic experiments. E. B. performed the zeolite synthesis and catalytic experiments. R.M. wrote the first manuscript draft with contributions from  R.G.-B. and M.B. All authors participated in the discussion and interpretation of the results, and contributed to the manuscript preparation and revision.

\section{Competing interests}
The authors declare no competing interests.

\clearpage 

\begin{suppinfo}
\renewcommand\thefigure{S\arabic{figure}} 
\renewcommand\thescheme{S\arabic{scheme}}
\renewcommand\thetable{S\arabic{table}}  
\setcounter{figure}{0}
\setcounter{scheme}{0}
\setcounter{table}{0}
\setcounter{page}{2}
\maketitle

\begin{table}[H]
\centering   
\begin{tabulary}{\textwidth}{   C   | C | C | C | C | C | C | C }
\hline
\rowcolor[HTML]{F4F4F4}   
\textbf{Formulas} & \textbf{Si/Al} & \textbf{Al}  & \textbf{Si}  & \textbf{\ce{[Cu(NH3)2]+}} & \textbf{\ce{NH4+}} & \textbf{\ce{NH3}} & \textbf{\ce{H+}} \\ \hline

\ce{H2Al2O192Si94}  & 47 & 2 & 94 & 0 & 0 & 0 & 2 \\ \hline
\ce{H8Al2N2O192Si94} & 47 & 2 & 94 & 0 & 2 & 0 & 0 \\ \hline
\ce{H12Al3N3O192Si93} & 31 & 3 & 93 & 0 & 3 & 0 & 0 \\ \hline
\ce{H15Al3N4O192Si93} & 31 & 3 & 93 & 0 & 3 & 1 & 0 \\ \hline
\ce{H18Al3N5O192Si93} & 31 & 3 & 93 & 0 & 3 & 2 & 0 \\ \hline
\ce{H28Al7N7O192Si89} & 12.7 & 7 & 89 & 0 & 7 & 0 & 0 \\ \hline
\ce{H12Al2Cu2N4O192Si94} & 47 & 2 & 94 & 2 & 0 & 0 & 0 \\ \hline
\ce{H10Al2Cu1N3O192Si94} & 47 & 2 & 94 & 1 & 1 & 0 & 0 \\ \hline
\ce{H18Al2Cu2N6O192Si94} & 47 & 2 & 94 & 2 & 0 & 2 & 0 \\ \hline
\ce{H18Al3Cu2N6O192Si93} & 31 & 3 & 93 & 2 & 1 & 0 & 0 \\ \hline
\end{tabulary}
\captionsetup{width=\textwidth}
\caption{Chemical composition, cationic species and molecules included in the triclinic \ce{T96O192} supercell models used for active learning and adversarial attack.}
\label{tab:dataset_composition_training}
\end{table}

\begin{table}[H]
\centering   
\begin{tabulary}{\textwidth}{   C   | C | C  }
\hline
\rowcolor[HTML]{F4F4F4}   
\textbf{Al distribution} & \textbf{$\Delta G_\mathrm{act}$ (kcal/mol)} & \textbf{$\Delta G$ (kcal/mol)}  \\ \hline
SR1  & 3.9 & 0.4 \\ \hline
SR2 & 4.6 & 1.7  \\ \hline
SR3 & 5.4 & 2.3  \\ \hline
SR4 & 4.1 & 0.5 \\ \hline
DR1 & 6.1 & 3.2  \\ \hline
DR2 & 6.7 & 2.9  \\ \hline
DR3 & 6.4 & 3.2  \\ \hline
DR4 & 5.4 & -0.7  \\ \hline
S4R & 7.3 & 4.5  \\ \hline
S6R & 6.4 & 5.4  \\ \hline
\end{tabulary}
\captionsetup{width=\textwidth}
\caption{Free energies of activation ($\Delta G_\mathrm{act}$) and reaction ($\Delta G$)  for the diffusion of a \ce{[Cu(NH3)2]+} complex between two neighboring cages through 8R windows with different Al distribution obtained from NNP-based biased simulations at 423 K.}
\label{tab:free_energies_us_al_pairs}
\end{table}

\begin{table}[H]
\centering   
\resizebox{\textwidth}{!}{
\begin{tabulary}{\textwidth}{   c  |  c |  c  | c | c | c | c | c | c }
\hline
\rowcolor[HTML]{F4F4F4}   
\textbf{Name} & \textbf{Formulas} & \textbf{Si/Al} & \textbf{Al}  & \textbf{Si}  & \textbf{\ce{[Cu(NH3)2]+}} & \textbf{\ce{NH4+}} & \textbf{\ce{NH3}} & \textbf{\ce{H+}} \\ \hline

L20 &   \ce{H144Al26Cu20N46O1536Si742} &    28.5 & 26 & 742 & 20 & 6 & 0 & 0 \\ \hline
L4 &   \ce{H112Al26Cu4N30O1536Si742} &     28.5 & 26 & 742 & 4 & 22 & 0 & 0 \\ \hline
M20 &   \ce{H240Al50Cu20N70O1536Si718} &     14.3 & 50 & 718 & 20 & 30 & 0 & 0 \\ \hline
M4 &  \ce{H208Al50Cu4N54O1536Si718} &      14.3 & 50 & 718 & 4 & 46 & 0 & 0  \\ \hline
M20-H+ &  \ce{H150Al50Cu20N40O1536Si718} &  14.3 & 50 & 718 & 20 & 0 & 0 & 30  \\ \hline
M20-NH3 & \ce{H423Al50Cu20N131O1536Si718} & 14.3 & 50 & 718 & 20 & 30 & 60 & 0  \\ \hline
H6R &    \ce{H312Al68Cu20N88O1536Si700} &   10.3 & 68 & 700 & 20 & 48 & 0 & 0 \\ \hline
H8R &   \ce{H312Al68Cu20N88O1536Si700} &    10.3 & 68 & 700 & 20 & 48 & 0 & 0 \\ \hline
HR &    \ce{H312Al68Cu20N88O1536Si700} &    10.3 & 68 & 700 & 20 & 48 & 0 & 0 \\ \hline
HB &    \ce{H312Al68Cu20N88O1536Si700} &    10.3 & 68 & 700 & 20 & 48 & 0 & 0 \\ \hline

\end{tabulary}}
\captionsetup{width=\textwidth}
\caption{Chemical composition, cationic species and molecules included in the triclinic \ce{T768O1536} supercell models used for NNP-based regular molecular dynamics simulations. }
\label{tab:dataset_composition_MD}
\end{table}

\begin{table}[H]
\centering   
\begin{tabulary}{\textwidth}{  c  | c | C | C | C | C | C }
\hline
\rowcolor[HTML]{F4F4F4}   
\textbf{Sample}   &   \textbf{Si/Al} & \textbf{Al/\textit{cha} cage} & \textbf{Crystal size (nm)} & \textbf{BET surface area (m$^2$/g)}  & \textbf{Micropore area (m$^2$/g)} & \textbf{Micropore volume (cm$^3$/g)}\\ \hline

CHA07 & 7.3 & 1.4 & $\sim$100 & 562 & 532 & 0.26 \\ \hline
CHA13 & 12.7 & 0.9 & $\sim$400-800 & 556 & 551 & 0.27 \\ \hline
CHA23 & 23.3 & 0.5 & $\sim$100 & 550 & 490 & 0.25 \\ \hline
\end{tabulary}
\captionsetup{width=\textwidth}
\caption{Chemical composition and physico-chemical properties of the synthesized CHA zeolites.}
\label{tab:properties_synthesize_cha}
\end{table}

\begin{table}[H]
\centering   
\begin{tabulary}{\textwidth}{   C   | C | C | C | C  }
\hline
\rowcolor[HTML]{F4F4F4}   
\textbf{Sample} & \textbf{Si/Al} & \textbf{\%wt Cu} & \textbf{Cu/Al} & \textbf{Al/\textit{cha} cage} \\ \hline

CHA07\textunderscore1.5Cu & 7.3 & 1.58 & 0.11 & 0.16 \\ \hline
CHA07\textunderscore3.0Cu & 7.3 & 2.92 & 0.20 & 0.30 \\ \hline
CHA13\textunderscore1.5Cu & 12.7 & 1.55 & 0.18 & 0.17 \\ \hline
CHA23\textunderscore1.5Cu & 23.3 & 1.54 & 0.35 & 0.17 \\ \hline

\end{tabulary}
\captionsetup{width=\textwidth}
\caption{Chemical composition of the Cu-exchanged CHA zeolites.}
\label{tab:chemcomp_cu_exchange_cha}
\end{table}

\begin{figure}[H]
\captionsetup[subfigure]{labelformat=parens, justification=centering}
    \centering
        \begin{subfigure}[b]{0.4\textwidth}
            \centering
            \includegraphics[width=\textwidth]{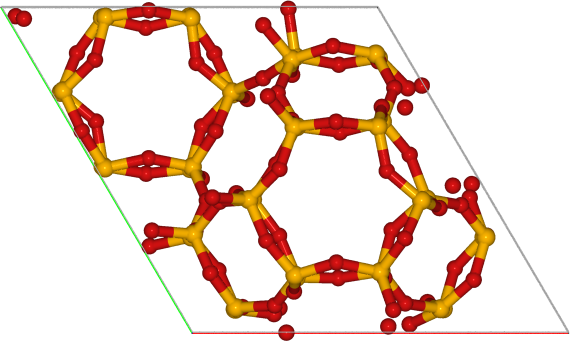}
        \caption{}
        \end{subfigure}
        \begin{subfigure}[b]{0.4\textwidth}
            \centering
            \includegraphics[width=\textwidth]{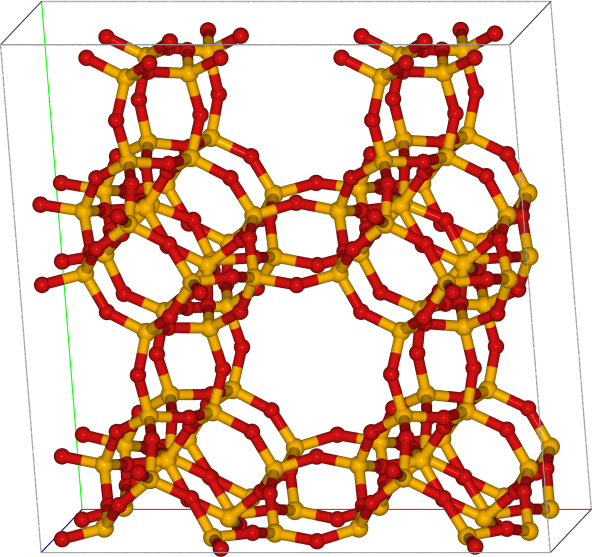}
        \caption{}
        \end{subfigure}
        \begin{subfigure}[b]{0.8\textwidth}
            \centering
            \includegraphics[width=\textwidth]{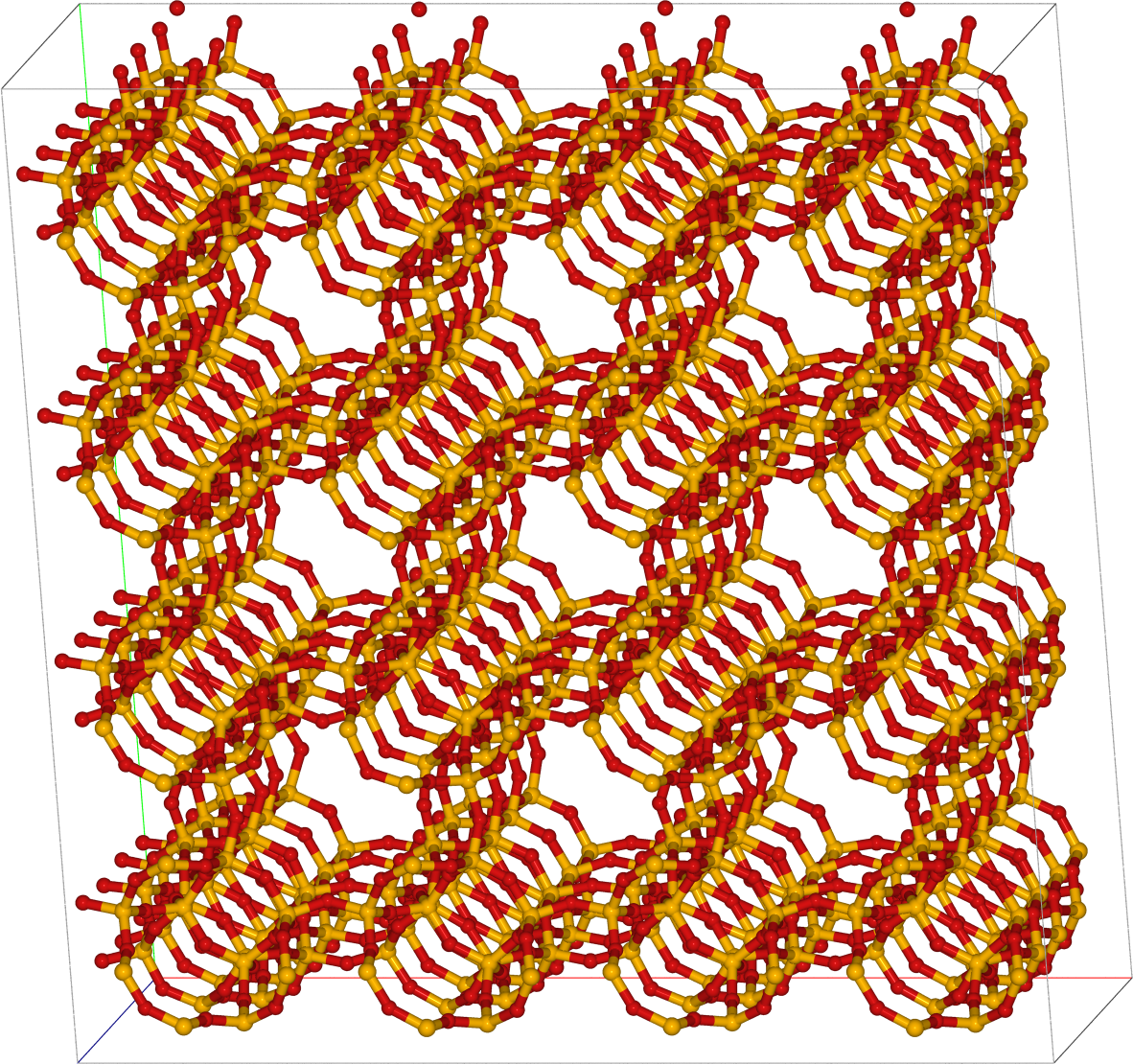}
            \caption{}
        \end{subfigure}
    \caption{Representation of the three types of base lattices used for umbrella sampling simulations (a and b) and for unbiased MD simulations (c). The representations correspond the all-silica CHA, however, the models used in this work include different combinations of Al number and locations starting from this base lattices, as explained in the main text. Color code: Si and O  are depicted as orange and red respectively.}
    \label{fig:snapshots_cell_sizes}
\end{figure}

\begin{figure}[H]
    \centering
    \includegraphics[width=0.6\textwidth]{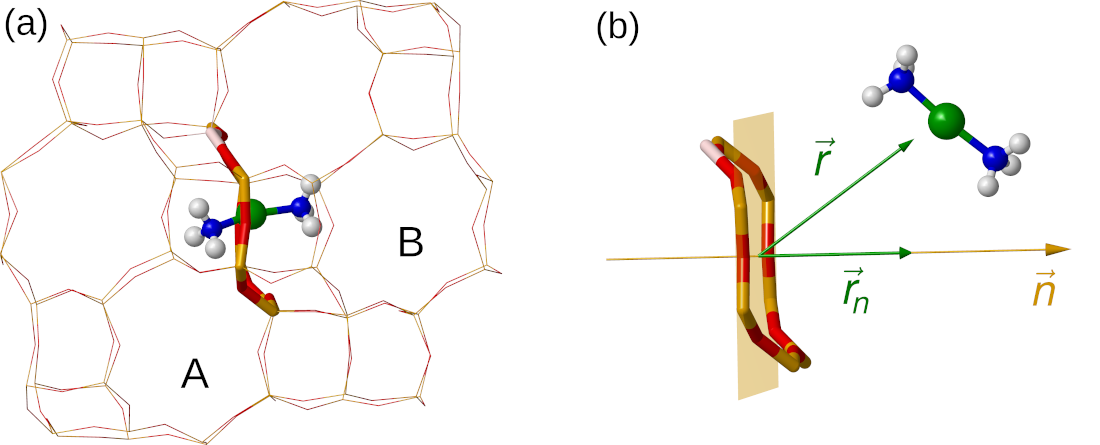}
   \caption{Diffusion of \ce{[Cu(NH3)2]+} complex from cage A ($\xi < 0$) to cage B ($\xi > 0$). (b) Representation of the collective variable $\xi$ describing the \ce{[Cu(NH3)2]+} diffusion through the 8R window (transparent yellow plane)}
    \label{fig:cv}
\end{figure}

\begin{figure}[H]
\raggedleft
\begin{subfigure}[b]{\textwidth}
    \includegraphics[width=\textwidth]{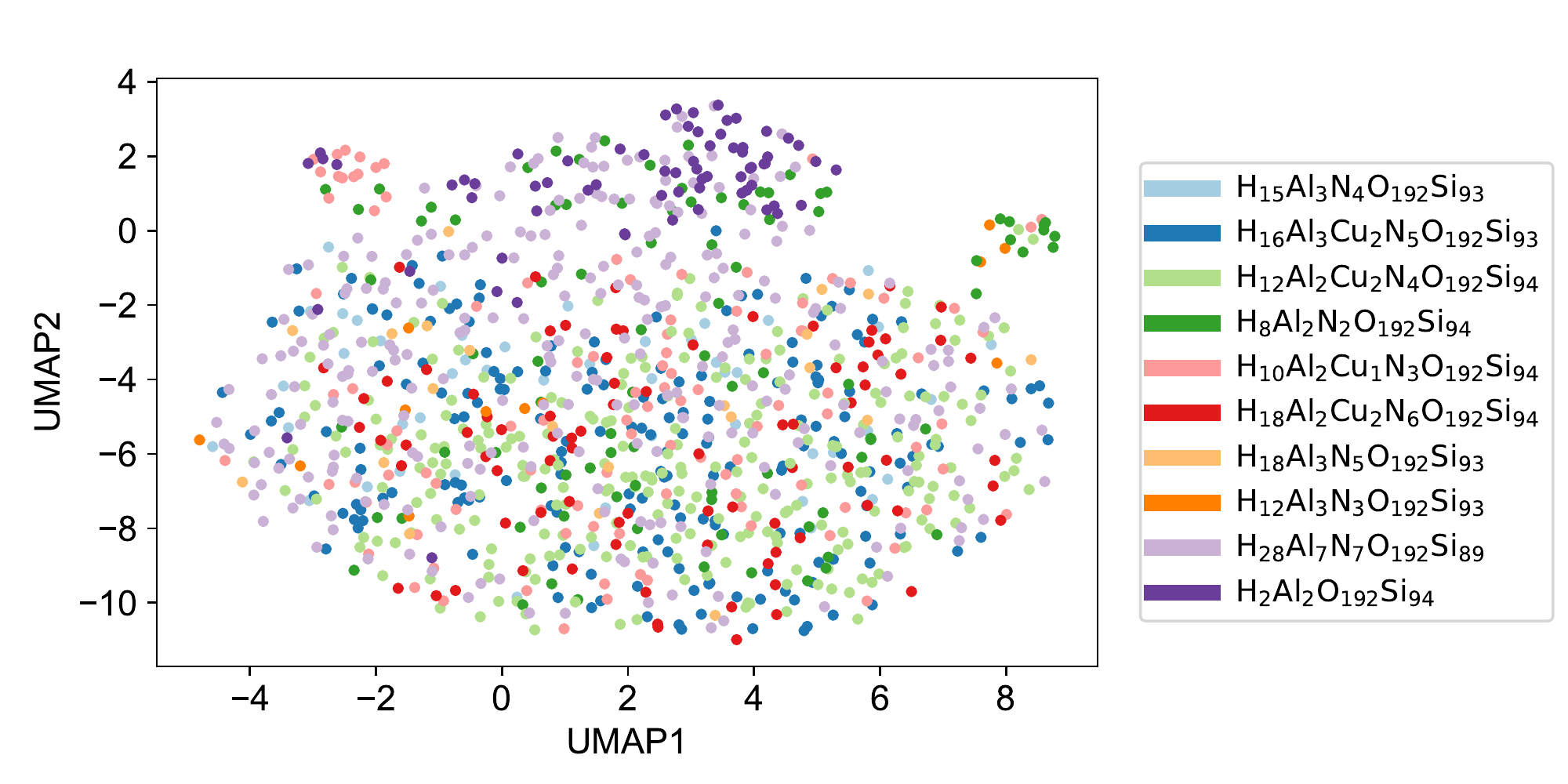}
\end{subfigure}
\caption{UMAP plot for the feature vectors of Al atoms from $\sim$1\% of the dataset. Both axes are on the same scale.}
\label{fig:features}
\end{figure}

\begin{figure}[H]
    \centering
    \begin{subfigure}[c]{\textwidth}
    \centering
   \includegraphics[width=\textwidth]{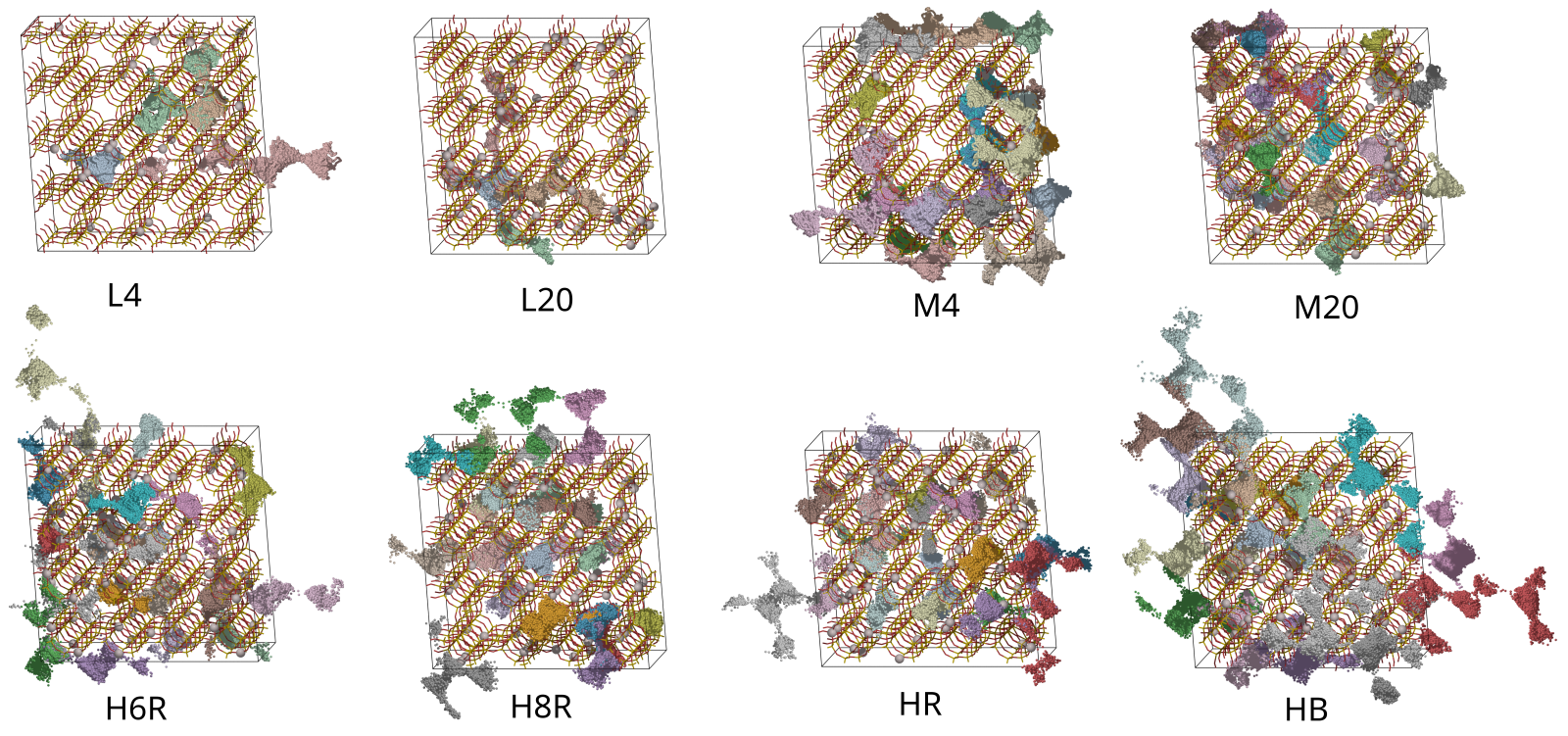}
    \end{subfigure}%
    \caption{Scatter plots showing the regions visited by \ce{Cu+} in the simulations run at 500 K for 5 ns. Color code: Si, O, and Al atoms are depicted as orange, red and light brown. Each \ce{Cu+} cation is represented with a random color.}
    \label{fig:scatterplots_Cu}
\end{figure}

\begin{figure}[H]
    \centering
    \begin{subfigure}[c]{\textwidth}
    \centering
     \includegraphics[width=\textwidth]{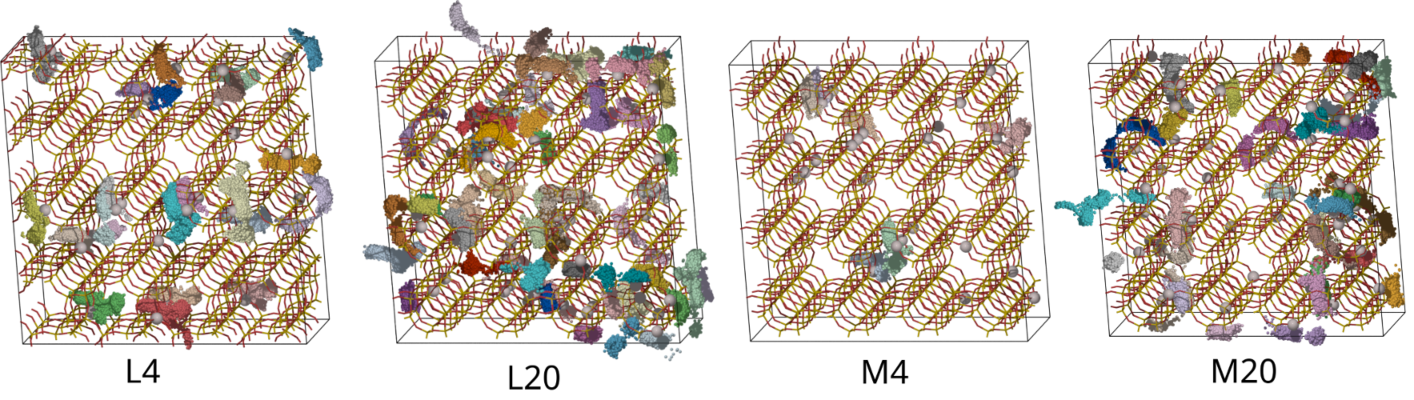}
    \end{subfigure}%
    \caption{Scatter plots showing the regions visited by \ce{NH4+} in the simulations run at 500 K for 5 ns. Color code: Si, O, and Al atoms are depicted as orange, red and light brown. Each \ce{NH4+} is represented with a random color.}
    \label{fig:scatterplots_nh4}
\end{figure}

\begin{figure}[H]
    \centering
    \includegraphics[width=\textwidth]{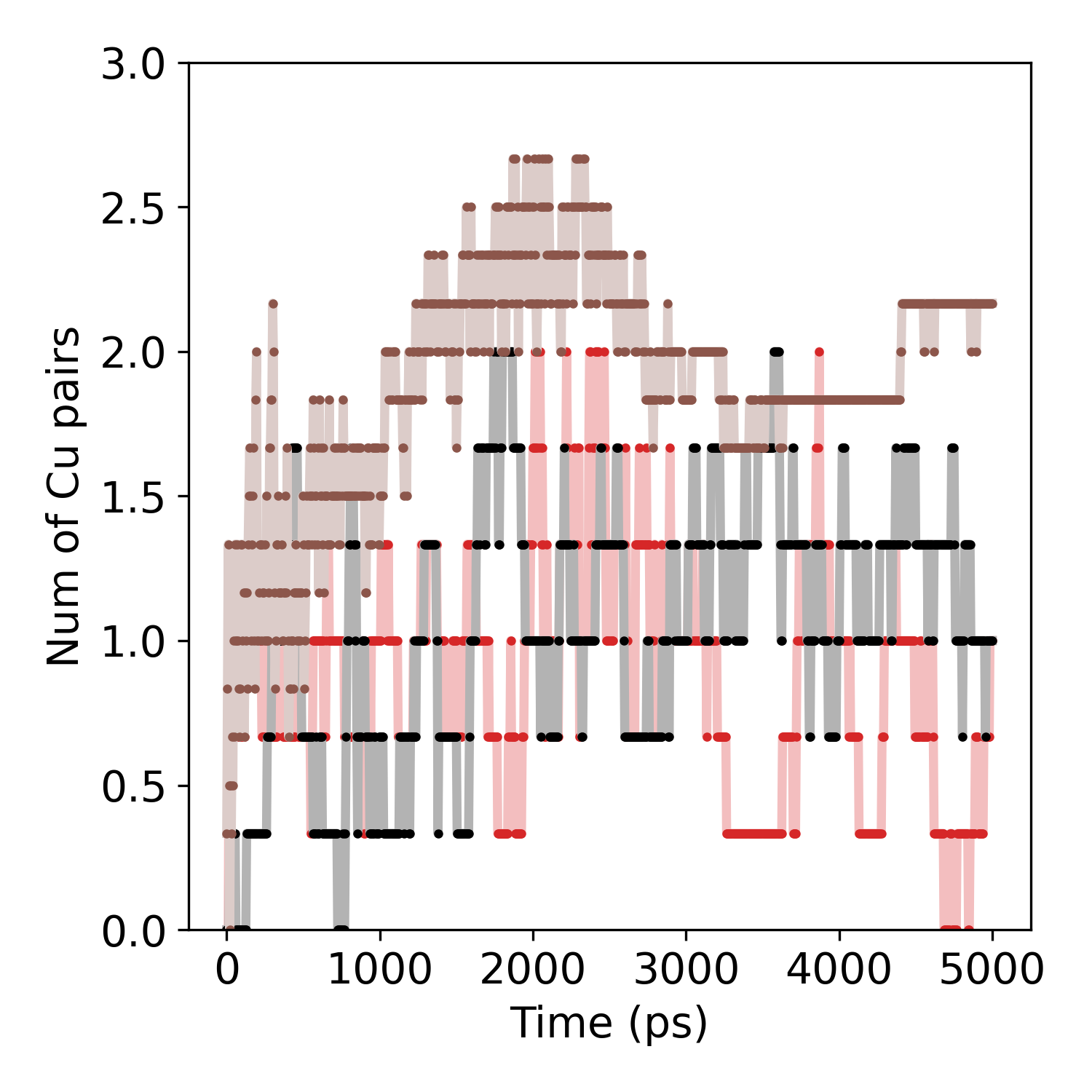}
    \caption{Time evolution of the average number of \ce{[Cu(NH3)2]+} pairs in the same cage for the models M20 (red), M20-H+ (black), M20-NH3 (brown), in unbiased MD simulations at 500 K.}
    \label{fig:cu_pairs_nh3}
\end{figure}

\begin{figure}[H]
    \centering
    \includegraphics[width=\textwidth]{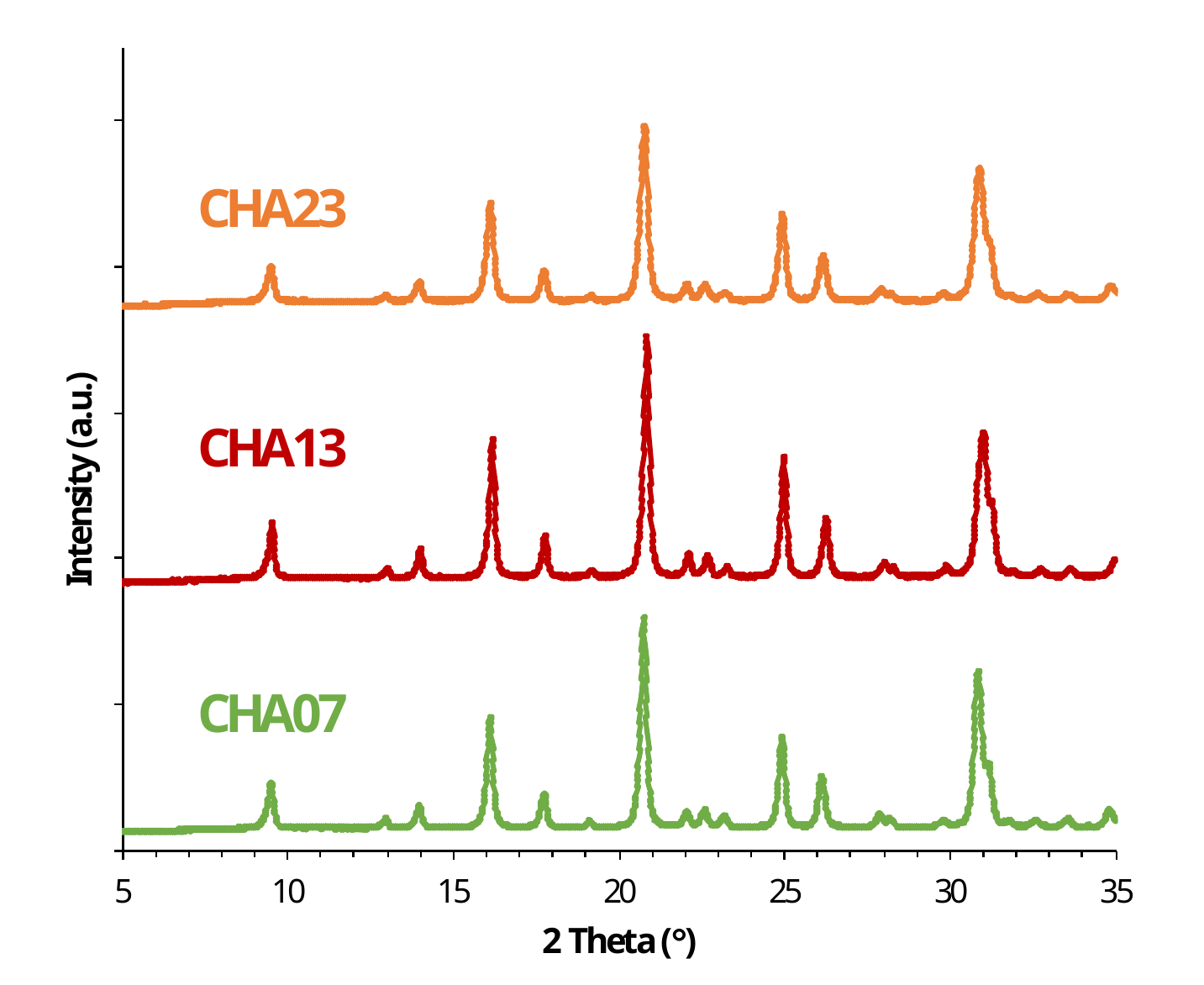}
    \caption{PXRD patters of the synthesized CHA zeolites}
    \label{fig:pxrd_patterns}
\end{figure}

\begin{figure}[H]
    \centering
    \includegraphics[width=\textwidth]{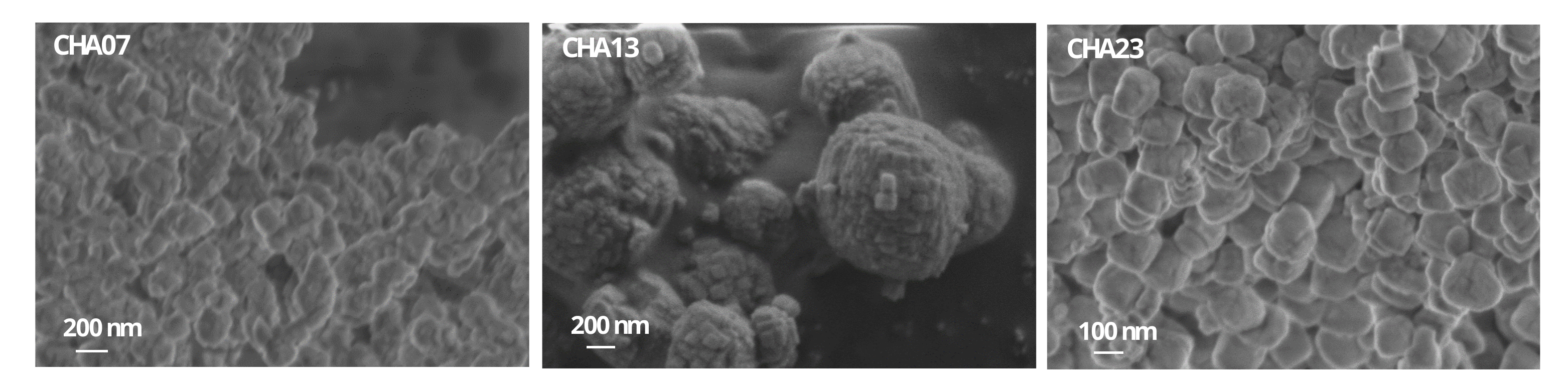}
    \caption{FESEM images of the synthesized CHA zeolites}
    \label{fig:fesem_images}
\end{figure}

\begin{scheme}[H]
    \centering
    \includegraphics[width=\textwidth]{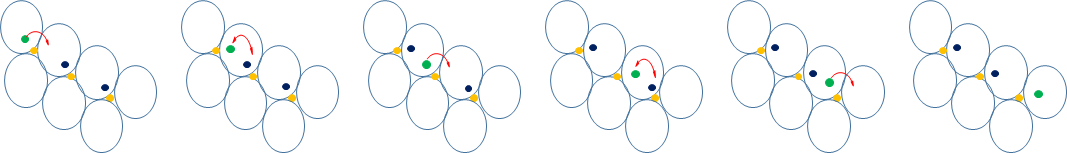}
    \caption{Representation of long-range \ce{[Cu(NH3)2]+} migration assisted by \ce{NH4+}. Al, \ce{Cu+} and \ce{NH4+} are represented with yellow, green and blue balls, respectively.}
    \label{fig:scheme_migration}
\end{scheme}

\end{suppinfo}

\end{document}